\documentclass[a4paper,11pt]{article}

\usepackage{amsmath,amssymb,bm,epsf,epsfig,graphicx,dsfont, float,multirow,tensor,url,placeins, subcaption}

\usepackage{hyperref}

\input epsf.sty
\numberwithin{equation}{section}
\numberwithin{figure}{section}

\def \rar {\rightarrow}

\def \la {\langle}
\def \ra {\rangle}
\def \rra {\rangle\!\rangle}

\def \ww {|\hspace{-1pt}|}
\def \ps {\phantom{+}}

\def \nn {\nonumber}
\def \Id {\mathds{1}}
\def \inf {\infty}

\newcommand{\re}{\mathop{\rm Re}\nolimits}
\newcommand{\im}{\mathop{\rm Im}\nolimits}

\def \pnn {\phantom{00}}
\def \pnnn {\phantom{000}}

\def\lS{\lambda_{\rm S}}
\def\lB{\lambda_{\rm B}}

\def \vv {{\mathcal V}}

\def \SU {{SU(2)\ }}
\def \SUk {{SU(2)$_k$\ }}
\def \SL {{SL(2,$\mathbb C$)\ }}

\def \dexp#1{\times 10^{#1}}

\newcommand{\dt}[1] {\frac{#1}{2}}

\begin{document}
\vskip 2.1cm

\centerline{\Large \bf Marginal deformations of SU(2)$_k$ WZW model boundary states }
\vspace*{4pt}
\centerline{\Large \bf  in open string field theory}
\vspace*{8.0ex}

\centerline{\large \rm Mat\v{e}j Kudrna\footnote{Email: {\tt matej.kudrna at email.cz}}}

\vspace*{4.0ex}
\begin{center}
{\it {FZU - Institute of Physics of the Czech Academy of Sciences,} \\
{Na Slovance 1999/2, 182 21 Prague 8, Czech Republic}}
\vskip .4cm

\end{center}
\vspace*{6.0ex}

\centerline{\bf Abstract}
\bigskip
We attempt to describe the moduli space of boundary states in the \SUk WZW model by constructing marginally deformed solutions in open string field theory in the level truncation approximation. In contrast with other approaches to marginal deformations, our solutions exhibit a $g$-function different from that of the background (typically lower). Thus, our method effectively combines features of both marginal and relevant deformations. After partially fixing an SU(2) symmetry of the equations of motion, we find families of solutions parameterized by the coefficient of the marginal field associated with the $J^3$ current, and we identify them as Cardy boundary states with varying angle $\theta$. However, it turns out that these solutions become inconsistent once the marginal parameter exceeds a certain value, implying that they cover only a part of the moduli space. Finally, we also compare the relation between the marginal parameter and the angle $\theta$ for different solutions and we find evidence suggesting that this relation is universal for certain classes of solutions.

\vfill \eject

\baselineskip=16pt

\tableofcontents

\setcounter{footnote}{0}

\section{Introduction and summary}\label{sec:intro}

In our previous work \cite{KudrnaWZW}, we investigated boundary states in the \SUk WZW model using open string field theory (OSFT) in the level truncation approach. We found many solutions that describe Cardy boundary states, which preserve the \SU symmetry of the model. However, the approach used in this reference produced only a finite number of isolated solutions and a discrete set of solutions obviously does not cover the moduli space of Cardy boundary states, which are parameterized by \SU group elements. In this work, we expand the previous results by combining the setup of \cite{KudrnaWZW} with the marginal deformation approach \cite{MarginalSen}. We have two main goals. First, we search for continuous families of solutions in the \SUk WZW model OSFT, and second, we analyze their properties in order to verify whether they cover the whole moduli space of boundary states.

Our main tool for investigating boundary states in the \SUk WZW model is Witten's open string field theory \cite{WittenSFT}, which provides a non-perturbative formulation of bosonic open string theory. For our purposes, it is important that there is a conjecture known as the background independence of OSFT \cite{SenBackground1}\cite{SenBackground4} (which is essentially proven by intertwining solutions \cite{ErlerMaccaferri}\cite{ErlerMaccaferri2}\cite{ErlerMaccaferri3}). This conjecture states that different D-brane backgrounds in OSFT are related by field redefinitions, which are realized by classical solutions of the OSFT equations of motion. Classical solutions are therefore conjectured to be in one to one correspondence with conformal boundary states. That means that by solving the OSFT equations of motion, it is possible to investigate boundary states in a given CFT and to find evidence about new boundary states \cite{KudrnaThesis}\cite{KudrnaWZW}.
Another conjecture, known as the Ellwood's conjecture \cite{EllwoodInvariants}\cite{KMS}, states that components of a boundary state corresponding to an OSFT solution are given by certain gauge invariant observables. This conjecture therefore provides a practical tool for assigning boundary states to OSFT solutions.

The level truncation approach \cite{SenZwiebachTV}\cite{RastelliZwiebach}\cite{MSZ lump}\cite{KMS}\cite{GaiottoRastelli}\cite{KudrnaUniversal}\cite{KudrnaThesis} is a numerical method in OSFT  based on truncating the string field, which has infinitely many degrees of freedom, to a finite set of states. The OSFT equations of motion are thus reduced to a system of polynomial equations, which can be solved using standard numerical methods.
Although the level truncation approach yields only approximate results, it nevertheless provides a great deal of valuable information. The thesis \cite{KudrnaThesis} presents numerous examples of OSFT solutions that were successfully identified as boundary states in several different models. Similarly, in \cite{KudrnaWZW}, we found many solutions that represent boundary states in the \SUk WZW model. Therefore, this method is a viable tool for investigating boundary states in a given BCFT.

The WZW models \cite{DiFrancesco}\cite{RecknagelSchomerus}\cite{Blumenhagen}\cite{WZWGoddard}\cite{WZWLectures}\cite{SchomerusLectures}\cite{WaltonWZWIntroduction} are one of the main groups of solvable conformal field theories. Their characteristic property is that they have an affine Lie algebra symmetry. In this work, we investigate boundary states in the simplest \SUk WZW model. Boundary states in this model can be divided into maximally symmetric boundary states, which preserve the \SU current symmetry, and into more generic symmetry-breaking boundary states, which have, in general, only the Virasoro symmetry. In \cite{KudrnaWZW}, we found OSFT solutions representing both types of boundary states. In this paper, we focus mainly on the maximally symmetric boundary states. These boundary states are well understood and they are given by the Cardy solution (\ref{Cardy BS}) in terms of \SU Ishibashi states. They are labeled by a half-integer label $J$, which goes from 0 to $k/2$, and by an \SU group element $g$. The \SU group manifold is isomorphic to a 3-sphere and therefore group elements $g$ can be parameterized by three angles $\theta$, $\psi$ and $\phi$.

In \cite{KudrnaWZW}, we imposed the condition
\begin{equation}\label{J03 psi int}
J_0^3 |\Psi\ra=0
\end{equation}
on the string field in order to fix a symmetry of the OSFT equations of motion and to simplify the problem. This condition imposes a strong restriction on the associated boundary states. It sets two of the three angles parameterizing $g$ to 0 and only a single angle $\theta \in (-\pi,\pi)$ remains. This is a great simplification because the \SU manifold can be now represented just by a circle instead of the 3-sphere.
Therefore we should have observed 1-parametric families of OSFT solutions, representing D-branes with varying $\theta$. However, we found only few discrete solutions on each background. By analyzing their invariants, we concluded that they represent boundary states with the angle $\theta$ following the rule
\begin{equation}\label{theta}
\theta=\pm(J_f-J_i+n)\frac{2\pi}{k},
\end{equation}
where $J_{i,f}$ are the initial and final values of the parameter $J$ and $n\in \mathbb{Z}$. The level truncation approximation therefore breaks the remaining part of the \SU symmetry and the solutions from \cite{KudrnaWZW} do not cover the moduli space of boundary states.

In this work, we attempt to restore the D-brane modulus by combining the approach from \cite{KudrnaWZW} with marginal deformations \cite{MarginalSen}\cite{MarginalKMOSY}\cite{MarginalTachyonKM}\cite{KudrnaThesis}\footnote{There are also several analytic solutions describing marginal deformations in OSFT, see for example \cite{MarginalSchnablPert}\cite{MarginalKiermaier}\cite{MarginalFuchs}\cite{MarginalKiermaierOkawa}\cite{MarginalMaccaferri}\cite{MarginalMaccaferriSchnabl}\cite{MarginalLaroccaMaccaferri}.}. Boundary spectrum of the \SUk WZW model includes three marginal fields associated to the three \SU currents. Two of them are removed by the condition (\ref{J03 psi int}) and only the field corresponding to the $J^3$ current remains. Therefore we search for families of OSFT solutions parameterized by the coefficient associated to this marginal field (which is denoted as $\lS$ to match the notation of \cite{MarginalTachyonKM}). These solutions are expected to represent boundary states connected by boundary marginal deformations induced by the $J^3$ current. An important difference between this paper and the references above is that our goal is to investigate solutions representing D-branes with $g$-function different from the background. Our solutions are therefore somewhat similar to the so-called vacuum branch of solutions mentioned in \cite{MarginalSen}\cite{MarginalKMOSY}, which is however considered to be nonphysical.

The search for a branch of solutions starts by setting $\lS=0$. For this $\lS$, the equations admit the solutions from \cite{KudrnaWZW} (which will be called basic or seed solutions), which represent Cardy branes with the angle $\theta$ given by (\ref{theta}). By turning on the marginal field, we can evolve the seed solutions into more generic solutions parameterized by $\lS$ that represent D-branes with generic values of the angle $\theta$. To make connection with the previous works, we parameterize $\theta$ as $\theta=\theta_0+\pi\lB$, where $\theta_0$ is the parameter of the seed solution and $\lB$ parameterizes the shift of the position of the D-brane with respect to the seed solution. Therefore there are two marginal parameters associated with our solutions, $\lS$ parameterizing the string field in OSFT and $\lB$ parameterizing boundary states in BCFT, and we can compare them like for other marginal solutions \cite{MarginalKMOSY}\cite{MarginalTachyonKM}\cite{KudrnaThesis}\cite{MarginalMaccaferriSchnabl}.

The families of solutions analyzed in this paper change the $g$-function of the background, which means that they are not purely marginal solutions, but they rather represent a combination of relevant and marginal deformations. From the boundary conformal field theory perspective, our results describe the following process: Consider some initial boundary state. This BCFT is first deformed by a relevant term of the form $\lambda\int \phi(x) dx$, where $\phi$ is a relevant primary field in the \SUk WZW model, which triggers an RG flow that stops at an RG fixed point $\lambda^\ast$ corresponding to a boundary state with a lower $g$-function. This is the boundary state associated to our seed solution. This boundary theory is then further deformed by a marginal term $\lambda_B\int J^3(x) dx$, which changes the $\theta$ parameter of the boundary state. Such boundary states correspond to solutions with nonzero $\lS$.

From a different perspective, using the background independence of OSFT, it as also possible to view the results as ordinary marginal deformations (i.e. preserving the $g$-function) of a \hbox{D-brane} represented by our seed solution, which are however described using degrees of freedom of a D-brane with different $g$-function (which corresponds to the initial background).

An interesting property of OSFT is that there are also solutions which describe boundary states with a $g$-function higher than the background \cite{Ising}\cite{KudrnaThesis}\cite{KudrnaWZW}. Such solutions describe irrelevant deformations of a boundary conformal field theory, which are difficult to realize using BCFT methods. Unfortunately, these solutions in the level truncation approximation often have low precision and other technical issues, like nonzero imaginary part at low levels, but at least some of them seem to be consistent. Therefore it is also possible to do marginal deformations of D-branes with higher $g$-function. One solution of this type will be discussed in section \ref{sec:k=3:B-brane}.

To summarize the results of this paper in one sentence, we can say that we have indeed found branches of solutions that consistently describe marginal deformations of the considered seed solutions, but they do not cover the whole moduli space of boundary states. To be more specific, let us consider the simplest case (which is discussed in section \ref{sec:k=2}): 0-brane boundary states in the $k=2$ model with $J=\frac{1}{2}$ initial boundary conditions. The branch of 0-brane solutions on this background has similar properties as the traditional marginal branch connected to the perturbative vacuum \cite{MarginalTachyonKM}\cite{KudrnaThesis}. It has a finite length (it becomes complex around $\lS\approx 0.33$), but it is effectively shorter because it goes off-shell and it becomes unphysical around $\lS^\ast\approx 0.16$.

\begin{figure}[!t]
   \centering
   \begin{subfigure}{0.35\textwidth}
   \includegraphics[width=\textwidth]{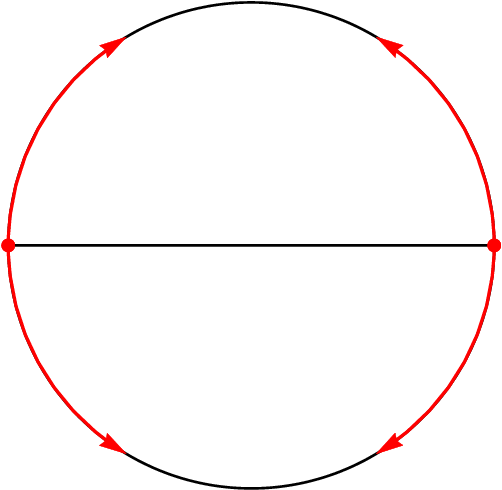}
   \end{subfigure}\qquad
   \begin{subfigure}{0.35\textwidth}
   \includegraphics[width=\textwidth]{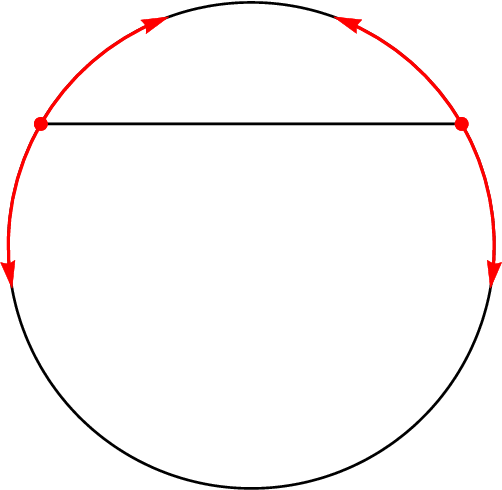}
   \end{subfigure}
   \includegraphics[width=0.35\textwidth]{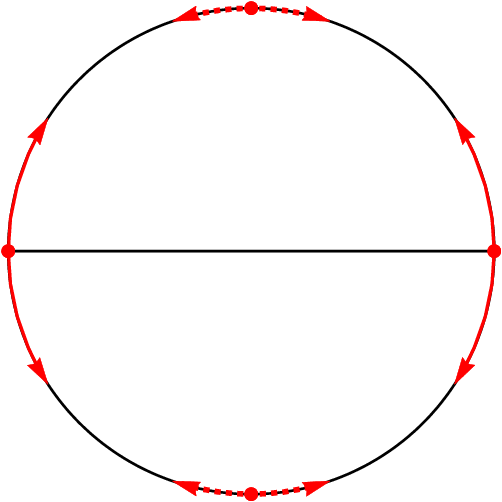}
   \caption{Visualizations of moduli space coverage by consistent 0-brane solutions in \SUk WZW models with $k=2$ (top left), $k=3$  (top right) and $k=4$ (bottom). The initial D-branes are denoted by black lines, the seed 0-brane solutions by red dots and the covered parts of moduli space by red arcs. The lengths of the arcs are based on the results from sections \ref{sec:k=2}, \ref{sec:k=3} and \ref{sec:k=4}. We were able to determine the maximal values of $\lB$ only with a limited precision, so the covered area may be somewhat larger or smaller. In case of one of the $k=4$ solutions (denoted by the dotted curves), we do not have a good estimate of maximal $\lB$, so this part of the figure is only schematic.}
   \label{fig:moduli}
\end{figure}

The first part of the branch is consistent with a 0-brane boundary state and the values of the boundary marginal parameter $\lB$ extracted from three different gauge invariants are almost identical. The marginal parameter goes up to approximately $\lB^\ast\approx 0.32$, while it has to reach at least $0.5$ to cover the whole moduli space. Therefore the on-shell part of the branch covers only approximately 64\% of the moduli space, see figure \ref{fig:moduli} for illustration. We are able to determine the maximal value of $\lB$ only with a limited precision, but it is obvious that the on-shell part of the branch does not cover the whole moduli space.

The second part of the branch seems to be inconsistent both as an OSFT solution (equations of motion that were removed during implementation of the level truncation scheme are violated in the infinite level limit) and as a boundary state (gauge invariants no longer match components of a 0-brane boundary state). Compared to the traditional marginal solution, the on-shell part of the branch is relatively shorter because only about one half of the branch is consistent.

Marginal deformations in models with higher $k$ have qualitatively the same properties. There are of course variations in lengths of branches and in precision of various observables, but the key features are mostly the same. The branches of solutions have finite length, they consistently describe marginally deformed boundary states in a limited range of $\lS$ and they cover only a part of the \SU moduli space.

The results of this paper suggest that the relations between the two marginal parameters $\lB$ and $\lS$ are not independent, but that they seem to be related for certain groups of solutions. In figure \ref{fig:lambda}, we plot the function $\lB(\sqrt{k}\lS)$ for several branches of solutions and the results are similar (for small enough $\lS$) for solutions that describe D-branes with the same final value of the parameter $J$, while the initial value seems to be irrelevant. The precision of our results is not good enough to say whether the agreement includes the whole function or just the leading term in its expansion, but it seems very likely that there are some relations between the functions.

Next, let us discuss how the coverage of the moduli space depends on the background. We noticed that the area covered by solutions connected to a single seed decreases with increasing level $k$. The difference in $\theta_0$ between two adjacent seed solution is $2\pi/k$, and since our results suggest that the areas around different seeds do not overlap, they naturally get smaller with increasing $k$. The initial boundary condition $J_i$ affects the coverage of moduli space through the number and positions of seed solutions. In \cite{KudrnaWZW}, we noticed an additional condition that (unless $J_i=k/2$) the seeds lie in the segment of the circle given by the initial D-brane, which implies that the remaining part of the circle without any seeds is left mostly uncovered.
Therefore the general trend is that the coverage of moduli space decreases with increasing $k$, especially for backgrounds which admit only a small number of seed solutions. A possible exception are backgrounds with $J_i=k/2$, which have seeds in the both halves of the circle.


It is unclear why on-shell solutions cover only a part of the moduli space and whether it is possible to describe the rest by some modification of our approach. We will suggest few speculations. One possibility is that there are other branches of solutions disconnected from the seed solutions which describe the missing parts of the moduli space (similarly to the toy model in \cite{MarginalTachyonKM}). But if such branches exist, they probably appear at higher levels and it will be very difficult to find them using the level truncation approach. A different possibility is that we observe only a part of the moduli space due to imposing Siegel gauge, which may not be compatible with all solutions. It may be necessary to consider a different gauge (or multiple gauges) to cover the rest of the moduli space. Another possibility is that there is an analytic solution for the whole family of boundary states, but this solution has some singular properties in parts of the moduli space, which do not allow its consistent truncation to a finite level. Finally, we cannot rule out the possibility that the observed inconsistencies could be caused by the level truncation approximation itself. It is possible that solutions can be interpreted as boundary states for any $\lS$, but the convergence towards the correct values for high $\lS$ becomes apparent only at a sufficiently high level (like 100 or 1000), which is currently unaccessible.

This paper is organized as follows. In section \ref{sec:setup}, we review some important formulas regarding the \SUk WZW model boundary states and OSFT observables, which will be needed later. In section \ref{sec:k=2}, we discuss marginal deformations of the basic 0-brane solution in the $k=2$ model. First, we analyze a sample solution with $\lS=0.05$ in detail and then we discuss properties of the full branch of solutions. Finally, we consider complex marginal deformations with purely imaginary value of the marginal parameter. In section \ref{sec:k=3}, we analyze marginal deformations in the $k=3$ model. We discuss the 0-brane solution and the symmetry-breaking B-brane solution. In section \ref{sec:k=4}, we perform a similar analysis of two 0-brane branches of solutions and one $\frac{1}{2}$-brane branch in the $k=4$ model. Finally, section \ref{sec:comparison} is dedicated to comparison of the relation between the two marginal parameters for various solutions.

\section{Preliminaries}\label{sec:setup}
The computational setup and execution of our calculations in this paper closely follow our previous work \cite{KudrnaWZW}, so we leave most of the technical details to this reference and to the thesis \cite{KudrnaThesis}, which provides a general description of our numerical algorithms. This section therefore mostly serves as a review of some important formulas regarding the \SUk WZW model boundary states and OSFT observables, which are needed for analysis of our solutions. We also briefly describe how we search for marginally deformed solutions, where we again mostly follow older references \cite{MarginalSen}\cite{MarginalKMOSY}\cite{MarginalTachyonKM}\cite{KudrnaThesis}.

\subsection{Cardy boundary states  in the \SUk WZW model}\label{sec:setup:SU2}
We will be working with the \SUk WZW model, where the level $k$ is a positive integer. Maximally symmetric boundary states in this model (which preserve one half of the bulk symmetry) satisfy gluing conditions of the form
\begin{equation}\label{Gluing J}
(J^a_n+\Omega^a_{\ b}(g)\bar J^b_{-n})\ww B\rra=0,
\end{equation}
where $\Omega(g)$ is the adjoint representation of an \SU group element $g$, which can be parameterized by three angles $\theta\in (0,\pi)$, $\psi\in (0,\pi)$ and $\phi\in (0,2\pi)$ as
\begin{equation}\label{SU2 group element}
g=\left(\begin{array}{cc}
\cos\theta+i \sin\theta \cos\psi & i e^{i \phi} \sin\theta \sin\psi \\
i e^{-i\phi} \sin\theta \sin\psi & \cos\theta-i \sin\theta \cos\psi
\end{array}\right).
\end{equation}

The structure of chiral primary states in the theory follows irreducible \SU representations and we denote them as $|j,m\ra$, where $j=0,1/2,\dots, k/2$ and $m=-j,\dots,j$. Notice that the number of representations that appear in a given \SUk WZW model is restricted by its level $k$.

The gluing condition (\ref{Gluing J}) is solved by Ishibashi states $|j,g\rra$, which are parameterized by the same half-integer label $j$ as primary states and a group element $g$. The lowest level components of Ishibashi states are
\begin{equation}\label{Ishibashi}
|j,g\rra=\sum_{m,n} (-1)^{j-m} D^j_{-mn}(g) |j,m,n\ra+\dots,
\end{equation}
where $|j,m,n\ra$ are bulk primary states and $D^j(g)$ is the Wigner D-matrix \cite{Hamermesh}.

Maximally symmetric boundary states are given by the Cardy solution. For each $g$, there is a set of boundary states
\begin{equation}\label{Cardy BS}
\ww J,g\rra=\sum_j B_J^{\ j}|j,g\rra=\sum_j\frac{S_J^{\ j}}{\sqrt{S_0^{\ j}}}|j,g\rra,
\end{equation}
where the label $J$ takes half-integer values $J=0,1/2,\dots, k/2$ and the modular $S$-matrix of the \SUk WZW model reads
\begin{equation}\label{S matrix}
S_{i}^{\ j}=\sqrt{\frac{2}{k+2}}\sin \frac{(2i+1)(2j+1)\pi}{k+2}.
\end{equation}
The boundary spectrum for any $g$ is given by irreducible \SU representations $|j,m\ra$, but the label $j$ is now an integer that goes from 0 to $\min(2J,k-2J)$.

In our work, we impose the following condition on the string field \cite{KudrnaWZW}\cite{Michishita}
\begin{equation}\label{J03 Psi}
J_0^3 |\Psi \ra=0.
\end{equation}
Its main purpose is to fix an exact symmetry of the OSFT equations $|\Psi\ra \rar e^{i\lambda_a J^a_0}|\Psi\ra$, which would prevent us from finding solutions using our numerical methods. This condition also significantly reduces the number of states in the \SUk WZW model Hilbert space, which speeds up the numerical calculations. Another consequence of (\ref{J03 Psi}) is that a string field which obeys this condition can describe only Cardy boundary states which satisfy
\begin{equation}
(J^3_n+\bar J^3_{-n})\ww B\rra=0.
\end{equation}
Group elements that are compatible with this condition are described just by a single angle $\theta\in (-\pi,\pi)$:
\begin{equation}
g=\left(\begin{array}{cc}
e^{i\theta} & 0 \\
0 & e^{-i\theta}
\end{array}\right).
\end{equation}
This leads to simplification of most formulas and the moduli space, which is originally isomorphic to a 3-sphere, is effectively reduced to a circle.

\subsection{Marginal deformations in OSFT}\label{sec:setup:marginal}
In the reference \cite{KudrnaWZW}, we solved OSFT equations in Siegel gauge and, after fixing the \SU symmetry using (\ref{J03 Psi}), we always obtained a discrete set of solutions. So we need to modify the numerical approach so that we can get continuous families of solutions which can potentially cover the whole moduli space.

First, we have to consider effects of the \SU symmetry $|\Psi\ra \rar e^{i\lambda_a J^a_0}|\Psi\ra$. That is an exact symmetry of the equations of motion because the zero modes $J_0^a$ are derivatives of the star product. It has formally 3 parameters, but only two of them are actually relevant because marginal solutions are always annihilated by a linear combination of the currents thanks to (\ref{J03 Psi}). Therefore this symmetry generates 2-spheres in the moduli space and the solutions from \cite{KudrnaWZW} actually cover two of the three dimensions of the moduli space. The main problem is therefore how to extend solutions in the remaining direction, which corresponds to the angle $\theta$ in our parameterization. The exact symmetry in this direction is broken by the level truncation approximation and a continuous family of solutions (of all Siegel gauge equations) does not exist. To deal with this problem, we follow the strategy of Sen and Zwiebach \cite{MarginalSen}, whose key point is that solutions are parameterized by a marginal field coefficient while one of the equations of motion is left unsolved.

The condition (\ref{J03 Psi}) eliminates two of the three marginal fields of the \SUk WZW model and only the marginal field given the $J^3$ current remains. We denote the coefficient of the marginal field as $\lS$ to match the notation from \cite{MarginalTachyonKM}\cite{KudrnaUniversal},
\begin{equation}\label{lambdaS def}
|\Psi\ra=\lS J_{-1}^3 c_1 |0,0\ra +\dots.
\end{equation}
Our goal is to find a continuous family of solutions which is connected to some initial solution. In \cite{KudrnaWZW}, we found that solutions describing Cardy boundary states (and most of other real solutions too) have vanishing value of the marginal field, so all initial solutions will have $\lS=0$ similarly to the perturbative vacuum, which serves as a seed for the traditional marginal solution.

In this approach, the coefficient $\lS$ is a parameter which can be set to an arbitrary value, so it can be no longer treated as a dynamical variable. Which means that we need to compensate for having one less dynamical variable and remove one equation in order to keep the equations of motion solvable. The canonical choice is the equation that corresponds to the marginal field
\begin{equation}
\la 0,0| c_{-1} J_1^3 |Q\Psi+\Psi\ast\Psi \ra.
\end{equation}
The remaining equations can be solved using Newton's method as usual. In practice, since we solve the equations of motion numerically, it is not possible to find the whole family of solutions and we compute only a finite number of sample solutions. Properties of the families of solutions change continuously with $\lS$, so a few samples is usually enough to understand their behavior.

Previous works regarding marginal deformations analyzed mainly the branch of solutions connected to the perturbative vacuum ($\Psi=0$). In this paper, we have a different goal and we will look for families of solutions which are connected to different boundary states at $\lS=0$, similarly to the 'vacuum' branch of solutions \cite{MarginalSen}\cite{MarginalKMOSY}. Technically, it means that the seed solution has a lower precision, but otherwise, we can proceed in the same way.

\subsection{OSFT Observables}\label{sec:setup:observables}
In order to identify OSFT solutions as boundary states, we consider several gauge invariant observables following \cite{KudrnaWZW}.

The first observable is the energy, which is defined using the OSFT action:
\begin{equation}\label{energy tot}
E_{tot}=E_J-S[\Psi]=E_J+\frac{1}{g_o^2}\left(\frac{1}{2}\la\Psi,Q\Psi\ra+\frac{1}{3}\la\Psi,\Psi\ast\Psi\ra\right).
\end{equation}
The normalization of the energy is chosen so that it correspond to boundary state $g$-functions. We take a $J$-brane with the group element $g=\Id$ as the background, which means the initial value of the energy $E_J$ equals to the $g$-function of the $J$-brane, $E_J=g_J=B_J^{\ 0}$.

In most OSFT calculations, the energy computed using the full action (\ref{energy tot}) equals to the energy computed just from the kinetic term
\begin{equation}\label{energy kin}
E_{kin}=E_J+\frac{1}{6 g_o^2}\la\Psi,Q\Psi\ra.
\end{equation}
However, this is no longer true if one of the equations of motion is left unsolved. The difference between the two expressions for the energy is then proportional to the omitted equation for the marginal field
\begin{equation}\label{energy dif}
E_{tot}-E_{kin}=\frac{1}{3 g_o^2}\lS \la 0,0|c_{-1}J_{1}^3 |Q\Psi+\Psi \ast\Psi \ra.
\end{equation}
The omitted equation should be equal to zero in the infinite level limit and both expressions for the energy should lead to the same result. Therefore we compute energy in both ways and we use that as a consistency check whether the equation (\ref{energy dif}) is asymptotically satisfied.

Next, we consider the so-called Ellwood invariants \cite{EllwoodInvariants}\cite{KMS}. The Ellwood invariants are observables based on on-shell bulk operators $\vv$ and they should reproduce the corresponding boundary state coefficients $\la \vv \ww B\rra$.
The condition (\ref{J03 Psi}) implies that Ellwood invariants of our solutions can be nonzero only for vertex operators $\vv$ that satisfy
\begin{equation}\label{Elw cond}
(J_0^3+\bar J_0^3)\vv=0.
\end{equation}

In this paper, we consider the same set of invariants as in \cite{KudrnaWZW}:
\begin{equation}\label{Elw def E}
E_{j,m} = 2\pi i\la E[c\bar c \phi_{j,m,-m}V^{aux}]|\Psi-\Psi_{TV}\ra,
\end{equation}
and
\begin{equation}\label{Elw def J}
J_{ab} = 2\pi i N_{ab}\la E[c\bar c J^a\bar{J^b}]|\Psi-\Psi_{TV}\ra.
\end{equation}
All invariants from the first set always satisfy (\ref{Elw cond}), while in the second set, there are only three potentially nonzero invariants: $J_{+-}$, $J_{-+}$ and $J_{33}$. We set their normalization to
\begin{eqnarray}
&&N_{+-}=N_{-+}=-\frac{1}{k},\\
&&N_{33}=-\frac{2}{k},
\end{eqnarray}
so that they match $E_{0,0}$ for universal solutions.

If one wants to restore the \SU symmetry fixed by (\ref{J03 Psi}), it is possible to act on solutions with operators $e^{i\lambda_1 J_0^1+i\lambda_2 J_0^2}$. These operators induce rotations on a 2-sphere, so the modified solutions will have more nonzero invariants and they will represent boundary states with nontrivial parameters $\psi$ and $\phi$. However, we will not do these calculations in this paper.

To identify solutions, it is necessary to know the expected values of our invariants for Cardy boundary states. By extracting the required boundary state components, we find that their absolute values are given by the matrix of boundary state coefficients $B_J^{\ j}$ and their phases by the angle $\theta$. For a boundary state with parameters $J$ and $\theta$, the expected values of Ellwood invariants are
\begin{eqnarray}\label{Elw inv exp}
E_{j,m}^{exp}&=&(-1)^{j-m}B_J^{\ j} e^{2 i m \theta}, \\
J_{\pm \mp}^{exp} &=& B_J^{\ 0} e^{\pm 2i \theta}, \\
J_{33}^{exp} &=& B_J^{\ 0}.
\end{eqnarray}

In this paper, we study marginal deformations of nontrivial solutions which describe Cardy branes with some (typically nonzero) initial angle $\theta_0$. Therefore we split $\theta$ into the initial angle $\theta_0$ and a term proportional to the boundary marginal parameter $\lB$:
\begin{equation}\label{theta lambda}
\theta=\theta_0+\Delta \theta=\theta_0+\pi \lB.
\end{equation}
To see why we added $\pi$ in front of $\lB$, let us consider marginal deformations of a boundary state in the BCFT formalism \cite{RecknagelSchomerusMarginal}
\begin{equation}\label{lambdaB def}
e^{2\pi i \lB J_0^3}\ww J,\theta\rra.
\end{equation}
The operator $e^{2\pi i \lB J_0^3}$ acts on primary states as
\begin{equation}
e^{2\pi i \lB J_0^3}|j,m,n\ra=e^{2\pi i \lB m}|j,m,n\ra.
\end{equation}
By comparison with (\ref{Ishibashi}), it follows that this deformation changes the angle as $\theta\rar \theta+\pi \lB$, which leads to (\ref{theta lambda}). The conventions for the marginal parameters are chosen to be similar to \cite{MarginalTachyonKM} and they guarantee that the relation between the marginal parameters is $\lB=\lS +O(\lS^3)$ for the standard marginal solution connected to the perturbative vacuum.

In order to obtain $\lB$ from a given invariant $E_{j,m}$, we have to invert the equation (\ref{Elw inv exp}). One possibility is treat the real part and the imaginary part separately:
\begin{eqnarray}
\lB &=& \frac{1}{2\pi m}\arccos \frac{(-1)^{j-m} \re[E_{j,m}]}{B_J^{\ j}}-\frac{\theta_0}{\pi}, \\
\lB &=& \frac{1}{2\pi m}\arcsin \frac{(-1)^{j-m} \im[E_{j,m}]}{B_J^{\ j}}-\frac{\theta_0}{\pi}.
\end{eqnarray}
However, this approach does not work very well because the numerical results do not reproduce the expected absolute values exactly and the two obtained values of $\lB$ are often inconsistent. A better option is to separate the invariant into its absolute value and a complex phase and to determine $\lB$ just from the complex phase:
\begin{equation}\label{lambda from elw E}
\lB = \frac{1}{2\pi m}\arg ((-1)^{j-m} E_{j,m})-\frac{\theta_0}{\pi}.
\end{equation}
$\lB$ can be also computed from the invariants $J_{\pm\mp}$. They lead to a similar expression for $\lB$:
\begin{equation}\label{lambda from elw J}
\lB = \pm\frac{1}{2\pi}\arg (J_{\pm\mp})-\frac{\theta_0}{\pi}.
\end{equation}

In addition to gauge invariant observables, we also compute the first 'out-of-Siegel' equation $\Delta_S$ \cite{KudrnaUniversal}\cite{KudrnaThesis} using the prescription
\begin{equation}\label{DeltaS}
\Delta_S=-\la 0| c_{-1}c_0 b_2|Q\Psi+\Psi\ast\Psi\ra.
\end{equation}
This quantity serves as a consistency check whether solutions satisfy one of the equations that were projected out during the implementation of Siegel gauge and it should approach zero.

Finally, let us make a brief comment on reality of solutions. In \cite{KudrnaWZW}, we showed how to derive reality conditions for components of the string field, but OSFT solutions in this model are often real only up to a null state $\chi$
\begin{equation}
\Psi^\ast=\Psi+\chi.
\end{equation}
Therefore it is better to determine reality of solutions from their observables. A real solution must satisfy
\begin{equation}\label{reality}
E_{j,m}=(-1)^{2j}E_{j,-m}^\ast
\end{equation}
and
\begin{equation}\label{reality2}
J_{+-}^\ast=J_{-+},\quad J_{33}^\ast=J_{33}.
\end{equation}

\section{0-brane solution in the $k=2$ model}\label{sec:k=2}
In this section, we consider the SU(2)$_2$ WZW model with $J=\dt1$ boundary conditions. As in \cite{KudrnaWZW}, we choose this setting to demonstrate properties of our solutions. It is the simplest nontrivial setting\footnote{The $k=1$ model is dual to the free boson theory on circle with the self-dual radius $R=1$ and it includes only 0-brane boundary states. Therefore it does not allow us to study transitions between different types of D-branes.} and therefore it allows us to reach the highest precision. Since we stored elements of the cubic vertex from the previous project, we can do OSFT calculations at the same level $L=14$ relatively easily because evaluation of the cubic vertex in this model requires significantly more computer resources than finding one solution.

At the beginning of our calculations, we set the marginal parameter to $\lS=0$. For this value of the marginal parameter, OSFT equations obviously admit the real 0-brane solution which was discussed in detail in \cite{KudrnaWZW}  (which is dual to one of the Ising model solutions \cite{Ising}\cite{KudrnaThesis}). This solution describes a 0-brane with $\theta_0=\frac{\pi}{2}$\footnote{There is also a solution with $\theta_0=-\frac{\pi}{2}$, which is related by a $Z_2$ symmetry, so there is no need to analyze it separately.}. By turning on the marginal field, we are able to find samples from the continuous family of solutions connected to this basic solution, which describe 0-brane boundary states with different values of $\theta$.

First, in subsection \ref{sec:k=2:example}, we choose one concrete value of the marginal parameter ($\lS=0.05$) and we analyze the corresponding solution in detail in order to illustrate properties of this type of solutions. Next, in subsection \ref{sec:k=2:lambda}, we discuss $\lS$-dependence of the 0-brane branch of solutions for real value of the marginal parameter in a similar way as in \cite{MarginalTachyonKM} and \cite{KudrnaThesis}. Finally, in subsection \ref{sec:k=2:complex}, we consider purely imaginary $\lS$, which leads to pseudo-real solutions that describe 0-brane boundary states in the \SL WZW model.

We also tried to find 0-brane solutions by an analogue of the tachyon approach from \cite{MarginalTachyonKM}, but it does not work in this model. If we fix the value of the tachyon field $t$ and remove the corresponding equation, there are no solutions with the desired interpretation. Depending on $t$, there are either no seeds with the required properties at level 2 or the promising seeds jump to different types of solutions when improved by Newton's method to levels 3 or 4. We suspect that this method does not work because the basic solution has a nonzero value of the tachyon field $t$ which changes with level.

\subsection{Example of a solution}\label{sec:k=2:example}
In this subsection, we analyze an example of a 0-brane solution for $\lS=0.05$ and compare its properties to the basic solution from \cite{KudrnaWZW} (see tables 4.1 and 4.2 in this reference), which has $\lS=0$.

We evaluated the 0-brane solution for $\lS=0.05$ up to level 14 and its independent observables are shown in table \ref{tab:sol 0.05}. The solution is real, so the remaining observables ($E_{1/2,-1/2}$, $E_{1,-1}$ and $J_{-+}$) follow from the reality conditions (\ref{reality}) and (\ref{reality2}).

\begin{table}[!]
\centering
\begin{tabular}{|l|llllll|}\hline
Level    & $E_{tot}$    & $E_{kin}$ & $E_{0,0}$ & $\ps E_{1,0}$  & $J_{33}$ & $\ps \Delta_S  $ \\\hline
2        & 0.752559     & 0.750197  & 0.73125   & $   -0.854816$ & 0.633958 & $\ps 0.01994539$ \\
3        & 0.741483     & 0.739718  & 0.722718  & $   -0.913018$ & 0.598321 & $\ps 0.00546991$ \\
4        & 0.728275     & 0.727052  & 0.720009  & $   -0.477370$ & 0.823577 & $\ps 0.00326187$ \\
5        & 0.725327     & 0.724258  & 0.717274  & $   -0.491410$ & 0.827218 & $\ps 0.00198586$ \\
6        & 0.720515     & 0.719661  & 0.714015  & $   -0.707192$ & 0.635631 & $\ps 0.00131989$ \\
7        & 0.719341     & 0.718559  & 0.712995  & $   -0.717529$ & 0.631307 & $\ps 0.00090972$ \\
8        & 0.716874     & 0.716212  & 0.712416  & $   -0.621105$ & 0.771000 & $\ps 0.00065135$ \\
9        & 0.716281     & 0.715660  & 0.711919  & $   -0.623477$ & 0.772206 & $\ps 0.00045064$ \\
10       & 0.714782     & 0.714238  & 0.710742  & $   -0.696016$ & 0.665352 & $\ps 0.00033709$ \\
11       & 0.714435     & 0.713918  & 0.710460  & $   -0.698382$ & 0.663886 & $\ps 0.00021891$ \\
12       & 0.713427     & 0.712964  & 0.710263  & $   -0.657977$ & 0.746139 & $\ps 0.00016680$ \\
13       & 0.713203     & 0.712760  & 0.710085  & $   -0.658775$ & 0.746603 & $\ps 0.00008944$ \\
14       & 0.712478     & 0.712075  & 0.709494  & $   -0.694758$ & 0.680308 & $\ps 0.00006647$ \\\hline
$\inf$   & 0.707103     & 0.707098  & 0.7065    & $   -0.706   $ & 0.708    & $   -0.000040  $ \\
$\sigma$ & $5\dexp{-9}$ & 0.000001  & 0.0003    & $\ps 0.007   $ & 0.031    & $\ps 0.000006  $ \\\hline
Exp.     & 0.707107     & 0.707107  & 0.707107  & $   -0.707107$ & 0.707107 & $\ps 0         $ \\\hline
\end{tabular}\vspace{5mm}
\begin{tabular}{|c|lll|}\hline
Level      & $\ps E_{1/2,1/2}$         & $\ps E_{1,1}$             & $\ps J_{+-}$              \\\hline
2          & $   -0.174708+0.729147 i$ & $   -0.682604-0.314159 i$ & $   -0.903462+0.314159 i$ \\
3          & $   -0.190140+0.741608 i$ & $   -0.660520-0.335501 i$ & $   -0.975216-0.401890 i$ \\
4          & $   -0.186838+0.751824 i$ & $   -0.652925-0.337513 i$ & $   -0.376612-0.462943 i$ \\
5          & $   -0.190883+0.772139 i$ & $   -0.645836-0.343348 i$ & $   -0.375174-0.132044 i$ \\
6          & $   -0.202108+0.776585 i$ & $   -0.636400-0.345703 i$ & $   -0.693300-0.138227 i$ \\
7          & $   -0.204397+0.779910 i$ & $   -0.632991-0.348763 i$ & $   -0.703262-0.359690 i$ \\
8          & $   -0.205170+0.782114 i$ & $   -0.630339-0.350279 i$ & $   -0.522928-0.375403 i$ \\
9          & $   -0.206462+0.787466 i$ & $   -0.628398-0.352549 i$ & $   -0.521151-0.246238 i$ \\
10         & $   -0.210532+0.788932 i$ & $   -0.624616-0.353662 i$ & $   -0.647330-0.250861 i$ \\
11         & $   -0.211433+0.790229 i$ & $   -0.623335-0.354864 i$ & $   -0.648888-0.350854 i$ \\
12         & $   -0.212009+0.791213 i$ & $   -0.622036-0.355681 i$ & $   -0.562521-0.357340 i$ \\
13         & $   -0.212637+0.793623 i$ & $   -0.621139-0.356856 i$ & $   -0.561539-0.290146 i$ \\
14         & $   -0.214753+0.794371 i$ & $   -0.619098-0.357500 i$ & $   -0.629120-0.293114 i$ \\\hline
$\inf$     & $   -0.2270\pnn+0.8081 i$ & $   -0.6040\pnn-0.3682 i$ & $   -0.603\pnnn-0.360  i$ \\
$\sigma$   & $\ps 0.0002\pnn+0.0001 i$ & $\ps 0.0003\pnn+0.0002 i$ & $\ps 0.025\pnnn+0.026  i$ \\\hline
$|\inf|$   & $\ps 0.8393             $ & $\ps 0.7074             $ & $\ps 0.702              $ \\
$|$Exp.$|$ & $\ps 0.840896           $ & $\ps 0.707107           $ & $\ps 0.707107           $ \\\hline
\end{tabular}
\caption{Gauge invariant observables of a 0-brane solution for $\lS=0.05$ in the $k=2$ model with $J=1/2$ boundary conditions. In the first part of the table, there are observables which should not depend on $\lS$, while invariants in the second part have nontrivial $\lS$-dependence. The line denoted $\inf$ represents infinite level extrapolations, the line denoted $\sigma$ their error estimates and the line denoted Exp. the expected values. In the second part of the table, there are additionally absolute values of extrapolations and their expected values.} \label{tab:sol 0.05}
\end{table}

In the first part of the table, there are invariants that should not depend on $\theta$. They include the energy, which should reproduce the 0-brane $g$-function (which is $1/\sqrt{2}$), the invariant $E_{0,0}$, which should also match the $g$-function, and invariants $E_{1,0}$ and $J_{33}$. We extrapolate these invariants using the procedure described in \cite{KudrnaThesis}. All extrapolations agree quite well with the values that are expected for a 0-brane. Compared to the original solution for $\lS=0$ from \cite{KudrnaWZW}, the precision is slightly lower, but this is an expected result because the equation for the marginal field is left unsolved. We observe that the two expressions for the energy $E_{tot}$ and $E_{kin}$ (based on (\ref{energy tot}) and (\ref{energy kin})) are close to each other, which suggests that this missing equation is solved at least asymptotically. The out-of-Siegel equation $\Delta_S$ is also satisfied well, so the solution seems to be consistent. Therefore the conclusion is that the basic OSFT properties of the solution are very similar to the solution for $\lS=0$ and that it behaves as a typical real Siegel gauge solution.

The table also includes estimated errors of extrapolations in the penultimate line. We observe that they over/underestimate the actual errors in similar way as in \cite{KudrnaWZW}, following a generic pattern discovered in \cite{KudrnaThesis}. Therefore we will not be concerned with these error estimates now and we will use them in a different way in the next subsection.

The second part of the table shows invariants which depend on $\theta$, which are needed for a full identification of the solution as a boundary state. These invariants have complex values, so it is useful to decompose them into absolute values and complex phases. Absolute values of extrapolations of the invariants are given in the penultimate row of the table. They serve mainly as a consistency check and we observe that their precision is comparable to other invariants.
The complex phases are more important because they are used to determine the BCFT marginal parameter $\lB$. We decompose $\theta$ as $\theta_0+\pi\lB$, where $\theta_0=\pi/2$ for the original 0-brane solution and, using (\ref{lambda from elw E}) and (\ref{lambda from elw J}), we obtain
\begin{eqnarray}
\lB^{(E_{1/2,1/2})} &=& 0.08716, \\
\lB^{(E_{1,1})}     &=& 0.08713, \\
\lB^{(J_{+-})}      &=& 0.086.
\end{eqnarray}
The first two values from $E_{1/2,1/2}$ and $E_{1,1}$ are very close to each other, the last value from $J_{+-}$ is less precise, because this invariant has much larger extrapolation error, but it is still consistent with the other two.
Therefore we conclude that this solution describes a consistent marginal deformation of the original solution with the BCFT marginal parameter equal approximately to
\begin{equation}
\lB\approx 0.08716.
\end{equation}

An interesting difference compared to the basic solution is that there are no longer any 'accidental' symmetries, which were typical for real solutions from \cite{KudrnaWZW}, so all gauge invariants which are not restricted by the reality condition are independent. This supports the conjecture from \cite{KudrnaWZW} that the 'accidental' symmetries are possible only when $\lS=0$ and that they are related to the duality to parafermion model solutions. In this particular model, when we turn off the marginal field, the basic solution is dual to one of Ising model solutions, which is built using the primaries of weight 0 and $\frac{1}{2}$. However, once we turn on the marginal field, which has no analogue in the Ising model because it has weight 1, the duality is broken and the 'accidental' symmetries disappear.

Another unusual property of real solutions mentioned in \cite{KudrnaWZW} is that invariants $E_{j,m}$ with $m$ close to $\pm j$ are more precise than invariants with $m$ around zero. For  families of marginally deformed solutions, this property gradually disappears with increasing value of $\lS$. This is not very apparent for this solution, which has low $\lS$ and only few invariants, but invariants of marginal solutions with the same weight have similar errors for high enough $\lS$.

\FloatBarrier
\subsection{$\lS$-dependence of the solution}\label{sec:k=2:lambda}
In this subsection, we investigate how properties of the marginally deformed 0-brane solution change with the marginal parameter $\lS$. We consider only positive $\lS$. The reason is that when we compare solutions with positive and negative $\lS$, all quantities are (anti)symmetric with respect to $\lS\rar -\lS$ and therefore it is enough to take $\lS\geq 0$.
Similarly to classical marginal deformations \cite{MarginalSen}, the branch has only a finite length and it is connected to complex solutions at higher $\lS$. The endpoint of the branch is approximately at $\lS\approx 0.33$ (it depends a bit on the level), but, as we are going to show below, solutions in the later part of the branch are no longer consistent. Therefore we will present data only up to $\lS=0.25$ with the interval between samples $\Delta \lS=0.01$.

\begin{figure}[!t]
   \centering
   \begin{subfigure}{0.45\textwidth}
   \includegraphics[width=\textwidth]{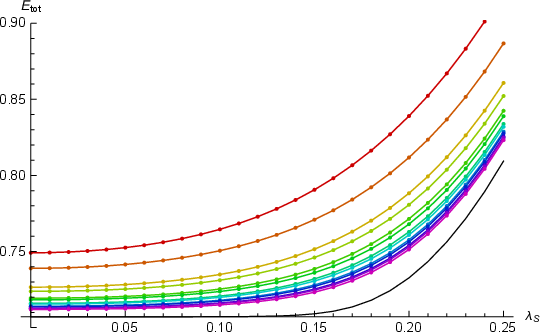}
   \end{subfigure}\qquad
   \begin{subfigure}{0.45\textwidth}
   \includegraphics[width=\textwidth]{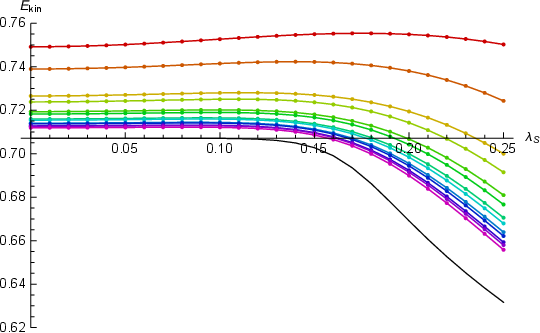}
   \end{subfigure}
   \vspace{3mm}

   \begin{subfigure}{0.45\textwidth}
   \includegraphics[width=\textwidth]{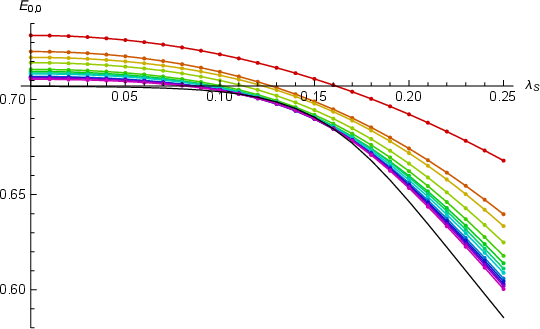}
   \end{subfigure}\qquad
   \begin{subfigure}{0.45\textwidth}
   \includegraphics[width=\textwidth]{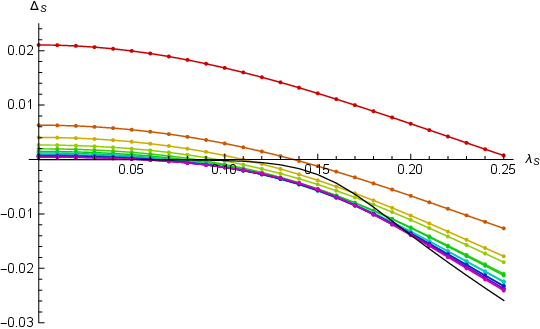}
   \end{subfigure}
   \caption{Plots of $\lS$-dependence of the energy (computed independently as $E_{tot}$, $E_{kin}$ and $E_{0,0}$) and the out-of-Siegel equations of the 0-brane branch of solutions in the $k=2$ model. Levels are distinguished by colors following the rainbow spectrum from red to purple and infinite level extrapolations are denoted by black lines.}
   \label{fig:k=2 energy}
\end{figure}

\begin{figure}[!t]
   \centering
   \begin{subfigure}{0.45\textwidth}
   \includegraphics[width=\textwidth]{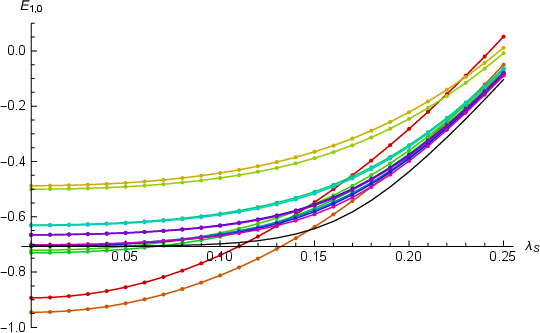}
   \end{subfigure}\qquad
   \begin{subfigure}{0.45\textwidth}
   \includegraphics[width=\textwidth]{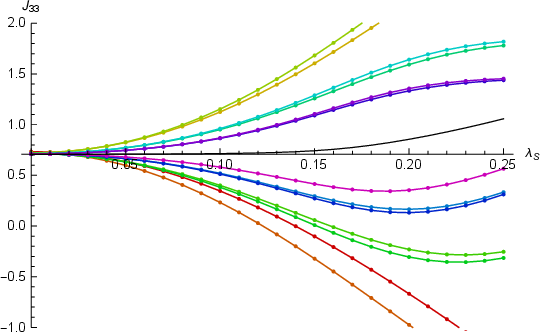}
   \end{subfigure}
   \vspace{3mm}

   \begin{subfigure}{0.45\textwidth}
   \includegraphics[width=\textwidth]{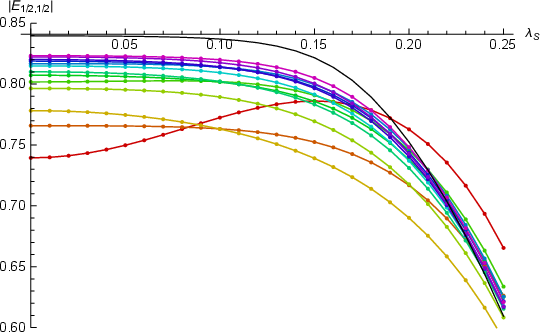}
   \end{subfigure}\qquad
   \begin{subfigure}{0.45\textwidth}
   \includegraphics[width=\textwidth]{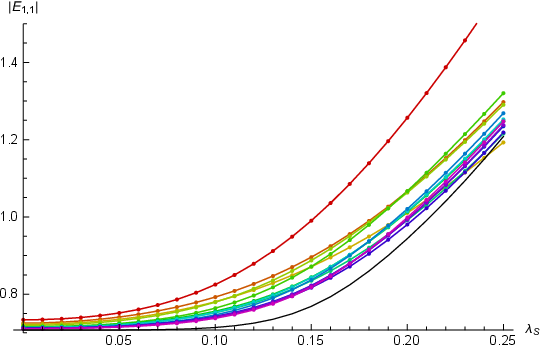}
   \end{subfigure}
   \caption{Plots of $\lS$-dependence of invariants $E_{1,0}$ and $J_{33}$ and absolute values of invariants $E_{1/2,1/2}$ and $E_{1,1}$ of the 0-brane solution in the $k=2$ model. The figures have the same style as in figure \ref{fig:k=2 energy}. These observables should be constant and their expected values are given by positions of the horizontal axes.}
   \label{fig:k=2 inv 1}
\end{figure}

\begin{figure}[!]
   \centering
   \begin{subfigure}{0.45\textwidth}
   \includegraphics[width=\textwidth]{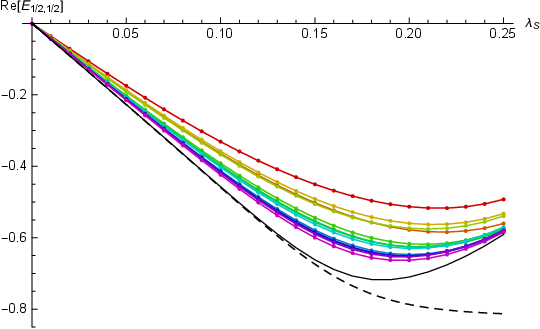}
   \end{subfigure}\qquad
   \begin{subfigure}{0.45\textwidth}
   \includegraphics[width=\textwidth]{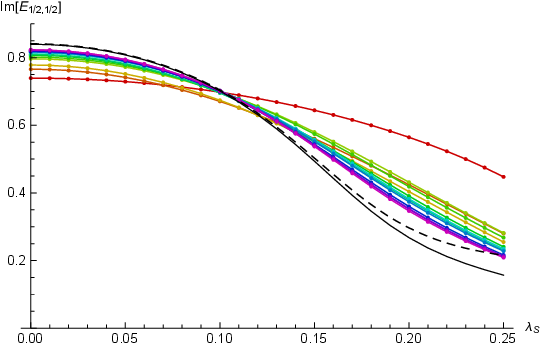}
   \end{subfigure}
   \vspace{5mm}

   \begin{subfigure}{0.45\textwidth}
   \includegraphics[width=\textwidth]{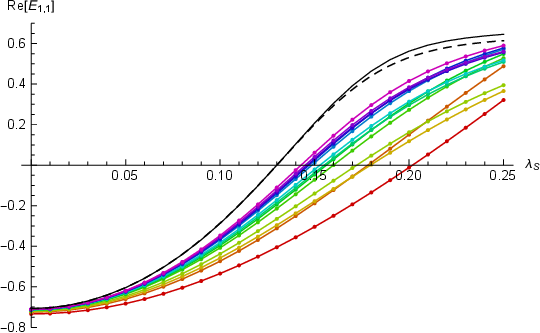}
   \end{subfigure}\qquad
   \begin{subfigure}{0.45\textwidth}
   \includegraphics[width=\textwidth]{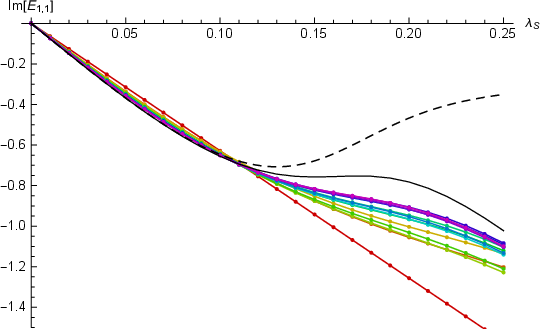}
   \end{subfigure}
   \vspace{5mm}

   \begin{subfigure}{0.45\textwidth}
   \includegraphics[width=\textwidth]{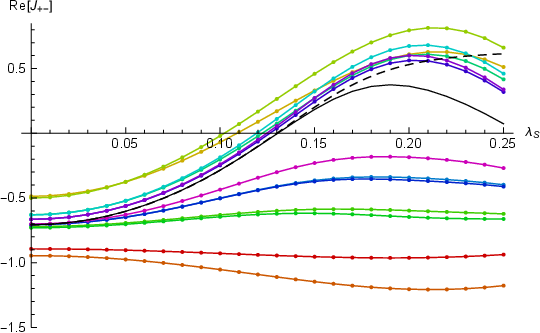}
   \end{subfigure}\qquad
   \begin{subfigure}{0.45\textwidth}
   \includegraphics[width=\textwidth]{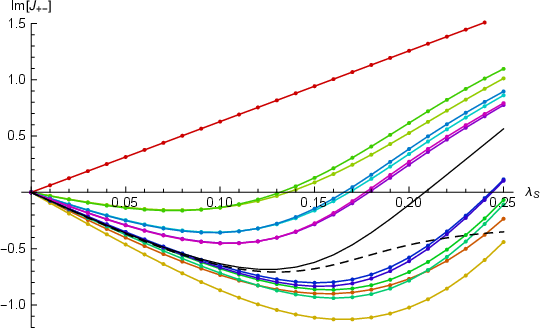}
   \end{subfigure}
   \caption{Plots of $\lS$-dependence of real and imaginary parts of invariants $E_{1/2,1/2}$, $E_{1,1}$ and $J_{+-}$ of the 0-brane solution in the $k=2$ model. The figures have the same style as in figure \ref{fig:k=2 energy}, the expected behavior (based on a fit of the phase of $E_{1/2,1/2}$ invariant) is denoted by dashed black lines.}
   \label{fig:k=2 inv 2}
\end{figure}

The dependence of various quantities on $\lS$ is plotted in figures \ref{fig:k=2 energy}, \ref{fig:k=2 inv 1} and \ref{fig:k=2 inv 2} in a similar way as in \cite{MarginalTachyonKM}\cite{KudrnaThesis}. The level is denoted by color following the rainbow spectrum from red at level 2 to violet at level 14. Infinite level extrapolations (using the method described in \cite{KudrnaThesis}) are denoted by solid black lines. The expected behavior is given either by the horizontal axes (for quantities that should be constant) or by dashed black lines (which are based on a fit of the phase of $E_{1/2,1/2}$ invariant, see later).

In the first figure, we plot the total energy $E_{tot}$, the energy from the kinetic term $E_{kin}$, the invariant $E_{0,0}$ and the out-of-Siegel equation $\Delta_S$. Since we know that this branch of solutions represents 0-brane boundary state, these quantities mainly serve as a consistency check. They should not depend on $\lS$ and they should be equal to the 0-brane $g$-function $g_{1/2}=2^{-1/2}$ ($E_{tot}$, $E_{kin}$, $E_{0,0}$) or to zero ($\Delta_S$). For small $\lS$, the curves representing infinite level extrapolations are almost constant and they pretty much coincide with the expected values. However, their behavior worsens as the value of $\lS$ increases and they gradually deviate from the expected values. We observe that at high enough $\lS$, three of the four quantities actually move away from the axis with increasing level and their infinite level extrapolations are even further away. Therefore it is highly unlikely that the energy of the solution converges to the expected 0-brane $g$-function for high $\lS$.

Next, figure \ref{fig:k=2 inv 1} includes invariants $E_{1,0}$ and $J_{33}$ and absolute values of invariants $E_{1/2,1/2}$ and $E_{1,1}$. All these quantities should be also independent on $\lS$. Once again, they behave according to our expectation for low $\lS$, but as the value of the marginal parameter increases, they start to deviate significantly from the expected values.

Therefore we think that a part of the branch is off-shell, similarly to the marginal branch in the traditional approach \cite{MarginalTachyonKM}. By saying that a solution is off-shell, we mean that it does not satisfy the full equations of motion and that its gauge invariants do not match components of a boundary state.
The figures in this subsection clearly illustrate that gauge invariants do not agree with a 0-brane boundary state components for high $\lS$, but there is still a small chance that a part of the branch could describe some symmetry-breaking boundary state. To eliminate this possibility, we do two additional consistency checks. In the level truncation approach in Siegel gauge, it is possible to solve only a subset of the full equations of motion\footnote{See \cite{ArroyoKudrna}\cite{KudrnaThesis} for a discussion of selection of equations to solve in different gauges.}, but the rest should be satisfied at least asymptotically. The first missing equation $\Delta_S$ (\ref{DeltaS}) is plotted in figure \ref{fig:k=2 energy} and we observe that it is violated for high $\lS$. We can also check the equation for the marginal field, which is proportional to the difference $E_{tot}-E_{kin}$, see (\ref{energy dif}). The two expressions for the energy converge to complectly different numbers for high $\lS$, so this equation is violated too.

An important question is where the branch of solutions goes off-shell. Answering this question is not easy because there is no sharp change in the behavior of observables and we can make only an approximate estimate.
One option how to investigate the problem is to use error estimates of extrapolations of various quantities. In \cite{KudrnaThesis}, we argued that this type of error estimates is not very reliable, so we cannot rely on the classical rules of statistical analysis, but there is a way how to use the error estimates. Let us take some observable $A$ and consider the ratio of its actual error (the difference between the infinite level extrapolation and the expected value) and its estimated error
\begin{equation}\label{sigma ratio}
\frac{\sigma}{\sigma_{est}}=\frac{A_{\inf}-A_{exp}}{\sigma_{est}}.
\end{equation}
For a single solution at given $\lS$, this ratio is not very meaningful because it typically takes quite diverse values for different observables. However, for a whole family of solutions it is possible to analyze its dependence on $\lS$. If we observe a quick growth of this ratio for multiple observables in some area, it is a good indication that the solution becomes inconsistent there.

\begin{figure}[!]
   \centering
   \includegraphics[width=10cm]{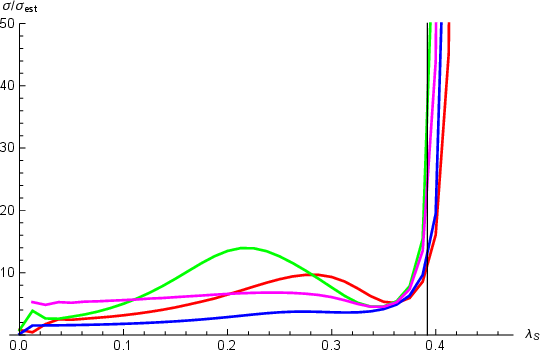}
   \caption{Plot of the ratio $\sigma/\sigma_{est}$ for several observables of the marginal branch solutions in the free boson theory (based on data from \cite{KudrnaThesis}). The four curves represent $E_{tot}$ (red), $E_{kin}$  (green), $E_{0}$ invariant (blue) and $\Delta_S$ (magenta). The black vertical line denotes the estimated point where the branch goes off-shell $\lS^\ast=0.392$.}
   \label{fig:sigma boson}
\vspace{20mm}
   \centering
   \includegraphics[width=10cm]{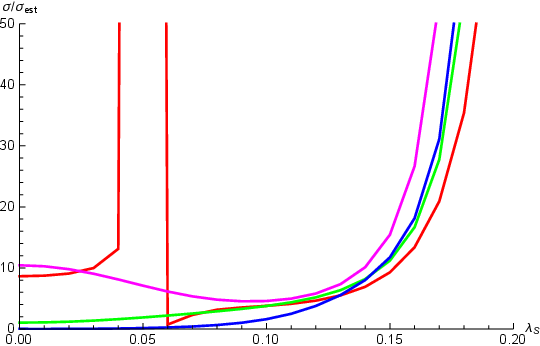}
   \caption{Plot of the ratio $\sigma/\sigma_{est}$ for several observables of the 0-brane solution in the SU(2)$_2$ WZW model. The curves represent $E_{tot}$ (red), $E_{0,0}$ invariant (green), $E_{1,0}$ invariant (blue) and $\Delta_S$ (magenta).}
   \label{fig:k=2 sigma}
\end{figure}

To test this approach, let us leave the current problem for a moment and consider the standard marginal deformations of the perturbative vacuum. We will refer to the free boson results \cite{MarginalTachyonKM}\cite{KudrnaThesis}, because they are known to the highest level, but the same solution can be found in the \SUk WZW model as well. For this marginal solution, there is an independent benchmark because we have an estimate of the end of the on-shell part of the marginal branch made using the tachyon approach \cite{KudrnaThesis}, $\lS^\ast\approx 0.392$.
The ratio (\ref{sigma ratio}) for several quantities of the free boson marginal solution is plotted in figure \ref{fig:sigma boson}. At first, we observe that the ratios take different values and that they do not change too much with $\lS$. However, as $\lS$ approaches the estimated endpoint (which is denoted by the vertical black line), all four ratios start to behave similarly and they grow very quickly. This indicates that this approach may give us a decent estimate of the endpoint.

Now, let us return to the SU(2)$_2$ WZW model. In figure \ref{fig:k=2 sigma}, we plot the ratio (\ref{sigma ratio}) for several observables of the 0-brane solution. Apart from the jump in the red curve (which is caused by $\sigma_{est}$ being coincidentally close to zero), we observe the same type of behavior as in figure \ref{fig:sigma boson}. The ratios are diverse for small $\lS$, but later, they start to grow quickly in a very similar manner as in the free boson case. The growth is somewhat slower, possibly due to lower level data. The comparison of figures \ref{fig:sigma boson} and \ref{fig:k=2 sigma} suggests that the branch of solutions goes off-shell slightly above $\lS=0.15$.

\begin{figure}[!t]
   \centering
   \includegraphics[width=10cm]{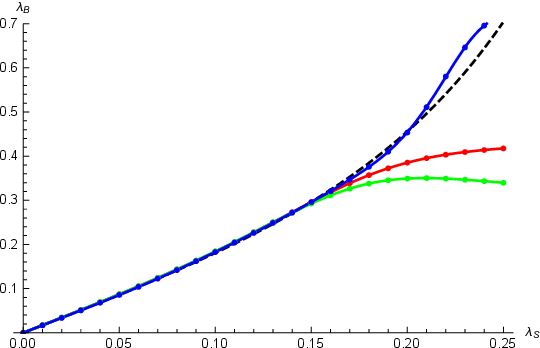}
   \caption{The relation between $\lS$ and $\lB$ for the 0-brane solution in the $k=2$ model. The three colors correspond to invariants $E_{1/2,1/2}$ (red), $E_{1,1}$ (green) and $J_{+-}$ (blue). The smooth functions are polynomial fits of the discrete data points. The dashed black line represents results from complex marginal deformations based on $E_{1/2,1/2}$ invariant, see section \ref{sec:k=2:complex}.}
   \label{fig:k=2 lambda}
\end{figure}

Next, we will focus on the boundary marginal parameter $\lB$, which measures the actual deformation of the basic 0-brane boundary state. Figure \ref{fig:k=2 inv 2} shows real and imaginary parts of invariants $E_{1/2,1/2}$, $E_{1,1}$ and $J_{+-}$. These invariants have a nontrivial dependence on $\lS$ and we use them to determine the parameter $\lB$ from their complex phases using (\ref{lambda from elw E}) and (\ref{lambda from elw J}). The resulting relations $\lB(\lS)$ for the three invariants are plotted in figure \ref{fig:k=2 lambda}.
We observe that the relation between the two parameters is linear for small $\lS$ as in the traditional approach, but the slope is different (the first coefficient in the expansion is approximately 1.7 instead of 1, see (\ref{lambda fits})).

In \cite{KudrnaThesis}, we were able to fit the data points with polynomials and use these fits to determine the first four coefficients in the expansion of $\lB(\lS)$ with decent precision. In our case, this approach does not work very well. The first few coefficients of polynomial fits of the three involved invariants are
\begin{eqnarray}\label{lambda fits}
\lB^{(E_{1/2,1/2})}(\lS)&=& 1.714 \lS + 11.4 \lS^3 + 124 \lS^5+\dots, \nn \\
\lB^{(E_{1,1})}    (\lS)&=& 1.712 \lS + 12.0 \lS^3 + 153 \lS^5+\dots,\\
\lB^{(J_{+-})}     (\lS)&=& 1.674 \lS + 15.4 \lS^3 + 111 \lS^5+\dots. \nn
\end{eqnarray}
The mutual agreement is not very good and it gets worse with increasing order. Therefore we are not able to extract a reliable parameterization of $\lB(\lS)$. We suspect that the difference compared to \cite{KudrnaThesis} lies in the basic solution. In the standard marginal approach, the starting point is the exact solution $\Psi=0$ at $\lS=0$, which means that the solution for small $\lS$ is very precise, while now we start from an approximate 0-brane solution, which reduces the precision for small $\lS$ and leads to differences in the fits.

In order to compare the fits of the invariants with the expected values, we computed order 10 fit of $\lB(\lS)$ based on the phase of the $E_{1/2,1/2}$ invariant and used it to indicate the expected values of the three invariants in figure \ref{fig:k=2 inv 2} by dashed black lines. For small $\lS$, the dashed lines agree quite well with the infinite level extrapolations, but they deviate later when the solution goes off-shell. That holds even for $E_{1/2,1/2}$ which is used for the prediction, because the absolute value of this invariant loses precision (see figure \ref{fig:k=2 inv 1}).

When we take a closer look at figure \ref{fig:k=2 lambda}, we observe that the values of $\lB$ from the three invariants are almost undistinguishable up to say $\lS=0.15$. For higher values, the three curves start to differ and it becomes clear that they do not correspond to a single relation. That is another sign of inconsistency of the solution. If we consider both the consistency of $\lB$ and the ratio of errors plotted in figure \ref{fig:k=2 sigma}, we make an educated guess\footnote{In order to get concrete numerical results, we consider the condition $\sigma/\sigma_{est}=X$, where $X$ is a number say in the interval $10\lesssim X\lesssim 30$ (an estimate based on figure \ref{fig:sigma boson}). A similar condition can be written for $\lB$, where we compare the results from different invariants to each other instead of the expected values. By varying the quantities we apply these conditions on and the number $X$, we get the presented average and error estimate.} that the branch goes off-shell at
\begin{equation}
\lS^\ast = 0.16\pm 0.01.
\end{equation}
That corresponds to following maximal value of the BCFT parameter
\begin{equation}\label{k=2 lB}
\lB^\ast=0.32\pm 0.02.
\end{equation}
The estimates may go a bit up or down depending on which invariants and parameters they are based on, but it is obvious that the truncated solution is consistent for small $\lS$ and inconsistent for high $\lS$.

In order to cover the whole moduli space, the maximal value of $\lB$ would have to be at least $1/2$. Our branch of solutions is consistent approximately up to $\lB\approx 0.32$, so it clearly does not cover the whole moduli space. The value (\ref{k=2 lB}) corresponds approximately to $64\%$ of the moduli space\footnote{This number corresponds to the percentage of covering the range of $\theta$. If we extend the solution by \SU rotations and consider the actual volume of the \SU manifold, the on-shell part of the branch covers approximately $93\%$ of volume of the 3-sphere.}, see figure \ref{fig:moduli} for graphical illustration.

Before we move on, let us return to figure \ref{fig:k=2 lambda}, which also includes a dashed black line which is based on the results from complex marginal deformations from the next subsection. There, we compute solutions with imaginary $\lS$ and we determine the relation between imaginary marginal parameters. To compare it with the results from the this subsection, the relation for imaginary parameters is fitted with a polynomial and evaluated it for real $\lS$. The presented curve is based on a 5-parametric fit of the invariant $E_{1/2,-1/2}$. It agrees well with the other curves up to $\lS^\ast$, which is a nice consistency check of our approach. The branch in the imaginary direction does not seem to go off-shell, so the results are, in principle, meaningful even above $\lS^\ast$. However, the transfer from imaginary to real parameter is sensitive to higher order terms in the fit, so the results are not reliable for much higher $\lS$ either. For $\lS>\lS^\ast$, the dashed curve is closest to the results from $J_{+-}$, which is somewhat surprising because this invariant has the lowest precision of the three invariants, but that can be just a coincidence.

\FloatBarrier
\subsection{Complex marginal deformations}\label{sec:k=2:complex}
In this subsection, we analyze marginal deformation solution with complex values of $\lS$, which describes 0-brane in the \SL WZW model. It is possible to consider generic complex $\lS$, but we will focus on purely imaginary $\lS$, which leads to pseudo-real\footnote{Pseudo-real solutions have some reality properties (some of their invariants and string field components are given by real numbers), but they do not satisfy the OSFT reality conditions (\ref{reality}) and (\ref{reality2}).} solutions. A similar branch of solutions in the classical marginal approach was briefly discussed in \cite{KudrnaThesis}.

The biggest difference compared to real solutions is that invariants that depend on $\theta$ either grow exponentially with the imaginary part of $\lB$ or they are exponentially suppressed because (\ref{Elw inv exp}) changes to
\begin{eqnarray}\label{Elw inv exp im}
E_{j,m}^{exp}&=&(-1)^{j-m}B_J^{\ j} e^{2 i m \theta_0}e^{-2\pi m\im [\lB]},\\
J_{\pm \mp}^{exp} &=& B_J^{\ 0} e^{\pm 2i \theta_0}e^{\mp 2\pi\im [\lB]}
\end{eqnarray}
for imaginary $\lB$. Therefore we also have to slightly modify the formula (\ref{lambda from elw E}) for extraction of $\lB$. For the invariants $E_{j,m}$, we find that
\begin{equation}\label{lambda from elw E im}
\im [\lB]=- \frac{1}{2\pi m}\log \frac{(-1)^{j+m}e^{-2 i m \theta_0}E_{j,m}}{B_J^{\ j}}
\end{equation}
and a similar formula holds for $J_{\pm\mp}$.

The branch of solutions with imaginary $\lS$ is again symmetric around the origin (although the symmetry is a bit different because it includes exchange of invariants $E_{j,m} \leftrightarrow E_{j,-m}$), so we consider only $\im [\lS]>0$. Similarly to the findings in \cite{KudrnaThesis}, the branch of solutions in the imaginary direction behaves better than for real $\lS$ and it continues longer. We tried tracking solutions up to $\lS=2i$ and it turns out that the length of the branch\footnote{In this case, the end of the branch means that the string field changes from pseudo-real to generically complex.} at individual levels is unpredictable and it oscillates a lot. For example, it ends slightly below $\lS=0.4i$ at level 2, around $0.82i$ at level 4, around $1.54i$ at level 5 and around $1.25i$ at level 6. At levels 3 and 7, the endpoints lie above $2i$. The only rule we noticed is that the branch tends to be longer at odd levels.

We evaluated sample solutions to the maximal available level 14 up to $\lS=0.4i$. That seems to be a good point to stop because the solution picks a generic complex part at level 2 around this point. The branch seems to be consistent in the whole considered range, see the analysis below, and it probably continues to be on-shell even for higher imaginary $\lS$, but the analysis gradually becomes impossible as more and more levels become fully complex and therefore unusable for extrapolations. Although we are not able to estimate the endpoint in the infinite level limit, it is clear that the on-shell part of the branch continues much longer than in the real direction, where it goes off-shell around $\lS \approx 0.16$.

\begin{figure}[!]
   \centering
   \begin{subfigure}{0.45\textwidth}
   \includegraphics[width=\textwidth]{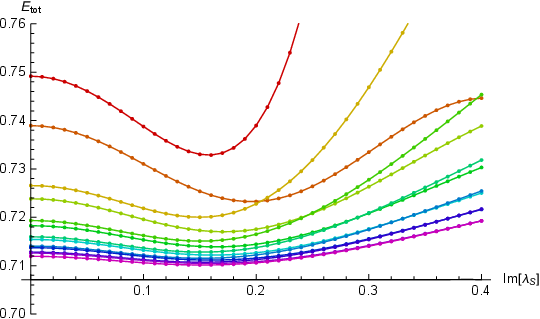}
   \end{subfigure}\qquad
   \begin{subfigure}{0.45\textwidth}
   \includegraphics[width=\textwidth]{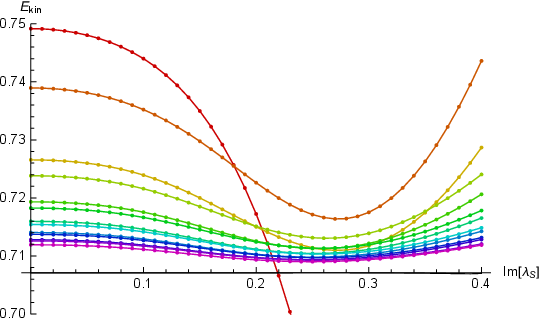}
   \end{subfigure}
   \vspace{5mm}

   \begin{subfigure}{0.45\textwidth}
   \includegraphics[width=\textwidth]{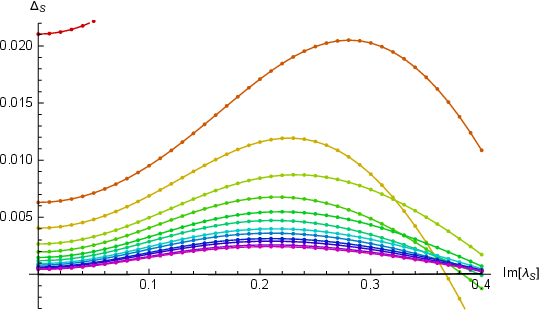}
   \end{subfigure}\qquad
   \begin{subfigure}{0.45\textwidth}
   \includegraphics[width=\textwidth]{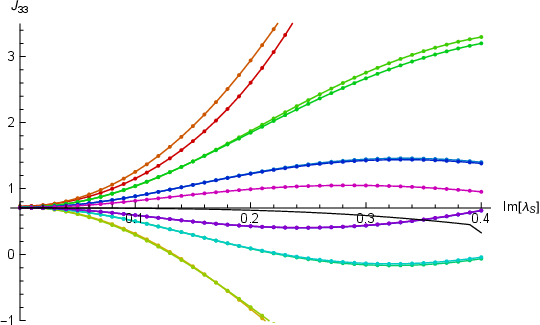}
   \end{subfigure}
   \vspace{5mm}

   \begin{subfigure}{0.45\textwidth}
   \includegraphics[width=\textwidth]{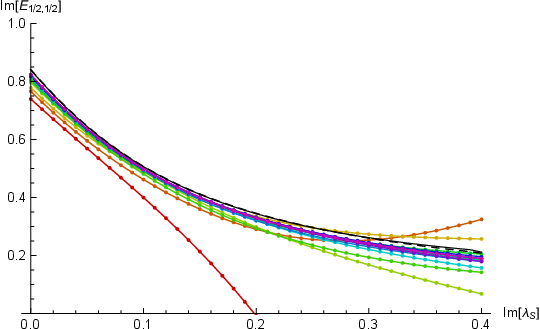}
   \end{subfigure}\qquad
   \begin{subfigure}{0.45\textwidth}
   \includegraphics[width=\textwidth]{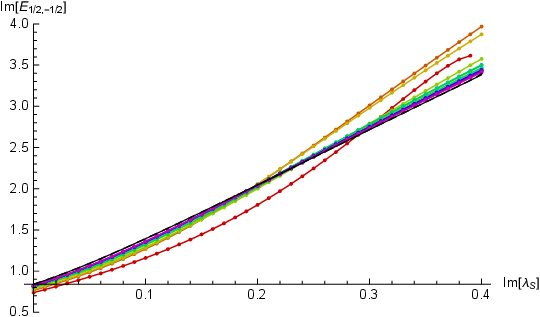}
   \end{subfigure}
   \vspace{5mm}

   \begin{subfigure}{0.45\textwidth}
   \includegraphics[width=\textwidth]{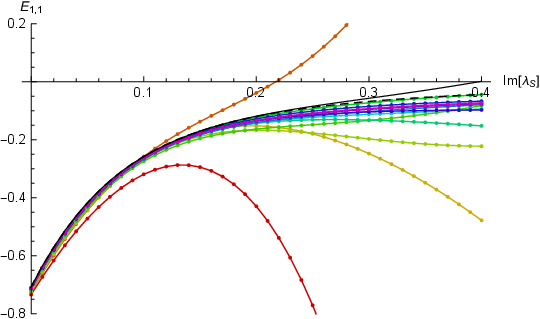}
   \end{subfigure}\qquad
   \begin{subfigure}{0.45\textwidth}
   \includegraphics[width=\textwidth]{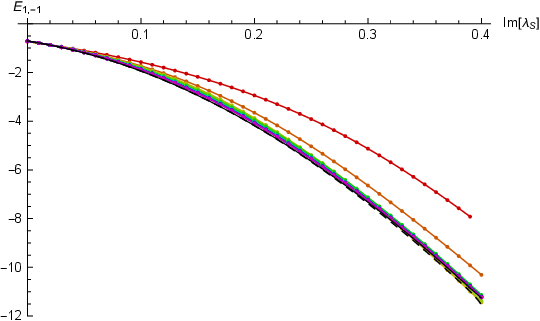}
   \end{subfigure}
   \caption{Plots of $\lS$-dependence of selected invariants for complex marginal deformations with imaginary $\lS$ of the 0-brane solution in the $k=2$ model.}
   \label{fig:k=2 complex}
\end{figure}

We show plots of $\lS$-dependence of selected observables in figure \ref{fig:k=2 complex}. First, let us check the energy. The total energy agrees very well with the predicted value $2^{-1/2}=0.707107$ in the whole considered range of $\im[\lS]$ and the infinite level extrapolation is almost indistinguishable from the axis. The same holds for $E_{kin}$. The missing equation for the marginal field is therefore satisfied well and the out-of-Siegel equation $\Delta_S$ too. That shows that the branch of solutions is most likely consistent at least up to $\lS=0.4i$.

An interesting property of this branch of solutions is that $E_{tot}$ has a local minimum at every level, which indicates that there is a solution of all Siegel gauge equations including the equation for the marginal field at every level. By approximating the solution at every level by a polynomial and evaluating it at the minimum, we checked that the minimum in fact corresponds to the \SL solution for 0-brane found in \cite{KudrnaWZW}. The positions of the minima jump up and down between even and odd levels and the reason is unclear. It is possible that the positions of the minima have no special meaning and they are determined just by the specific truncation of the string field at the given level. When we tried the same approach for models with higher $k$, we noticed that there are branches of solutions with no local minima at some levels. This is probably the source of the so-called odd level instabilities of some \SL solutions that were observed in \cite{KudrnaWZW}.

Although the branch of solutions is consistent in the whole considered range of $\lS$, there is obviously a gradual loss of precision. To illustrate that, there is a plot of the invariant $J_{33}$ in figure \ref{fig:k=2 complex}. Oscillations of the invariant grow with $\im [\lS]$, which leads to increase of error of its infinite level extrapolation.

Figure \ref{fig:k=2 complex} also includes invariants $E_{1/2,\pm 1/2}$ and $E_{1,\pm 1}$. Unlike in the real direction, these invariants are now either real or purely imaginary. However, invariants with the opposite sign of $m$ are no longer related by complex conjugation because they do not satisfy (\ref{reality}), so the number of independent observables remains the same.  These invariants depend nontrivially on $\lB$ and we observe that invariants with positive $m$ decrease in absolute value, while absolute values of invariants with negative $m$ grow quickly. The decrease or growth is exponential, which corresponds with the predicted behavior (\ref{Elw inv exp im}). This branch of solutions has a symmetry $E_{j,\pm m}(\lS)=E_{j,\mp m}(-\lS)$ when we switch the sign of $\im[\lS]$.

\begin{figure}[!t]
   \centering
   \includegraphics[width=10cm]{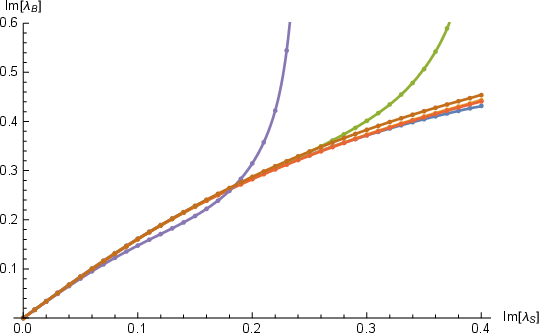}
   \caption{The relation between $\im[\lS]$ and $\im[\lB]$ for complex marginal deformations with imaginary $\lS$ of the 0-brane solution in the $k=2$ model. The results are based on invariants $E_{1/2,\pm 1/2}$, $E_{1,\pm 1}$ and $J_{\pm\mp}$. The two curves that deviate from the rest due to low numerical precision correspond to invariants $E_{1,1}$ (green) and $J_{+-}$ (purple).}
   \label{fig:k=2 complex lambda}
\end{figure}

Using (\ref{lambda from elw E im}), we computed the relation between $\lS$ and $\lB$ from the 6 invariants that depend on $\lB$. The results are plotted in figure \ref{fig:k=2 complex lambda}. Four of the six curves are close to each other, while the remaining two deviate from the rest. In this case, the deviation of the two curves does not indicate inconsistency of the solution, but it is just a result of loss of numerical precision. The two curves correspond to invariants $E_{1,1}$ (green) and $J_{+-}$ (purple). These two invariants decrease exponentially with $\lS$ (the plot of $E_{1,1}$ can be seen in the bottom left part of figure \ref{fig:k=2 complex}) while errors of their extrapolations grow. At some point, their relative errors inevitably exceed 100\%, which obviously means that the extracted values of $\lB$ become completely unreliable, which causes the deviation of the two curves. The other four invariants have much smaller relative errors and the extracted values of $\lB$ agree quite well, so we believe that the solution is consistent in the whole considered range of $\lS$.

In order to compare results for real and imaginary $\lS$, we fitted the data from $E_{1/2,-1/2}$ by a polynomial, evaluated the fit for real $\lS$ and plotted it in figure \ref{fig:k=2 lambda}. It agrees well with the results for real $\lS$, which suggests that all our numerical solutions are approximations of a single solution analytic in $\lS$. This exact solution probably has
poles or branch cuts on the real axis, which limits the length of the branch in the real direction.

\FloatBarrier
\section{Solutions in the $k=3$ model}\label{sec:k=3}
In this section, we discuss marginal deformations of solutions in the $k=3$ model with $J=1/2$ boundary conditions. In \cite{KudrnaWZW}, we found two real solutions this model, a 0-brane and a B-brane (which is a symmetry-breaking boundary state \cite{WZW B-branes}). We analyze marginal deformations of these solutions up to level 12. Many properties of solutions in the $k=3$ model are similar to the $k=2$ model, so our analysis will be more brief and we will focus on differences between $k=2$ and $k=3$ solutions.

\subsection{0-brane solution}\label{sec:k=3:0-brane}
\begin{figure}[!t]
   \centering
   \begin{subfigure}{0.45\textwidth}
   \includegraphics[width=\textwidth]{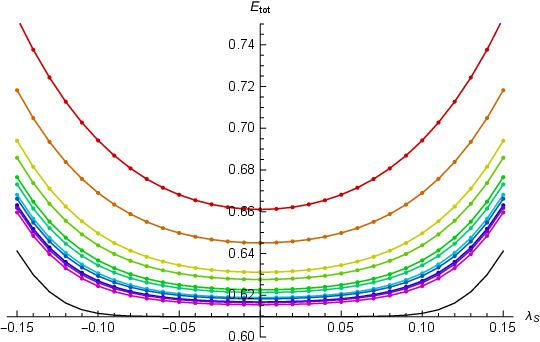}
   \end{subfigure}\qquad
   \begin{subfigure}{0.45\textwidth}
   \includegraphics[width=\textwidth]{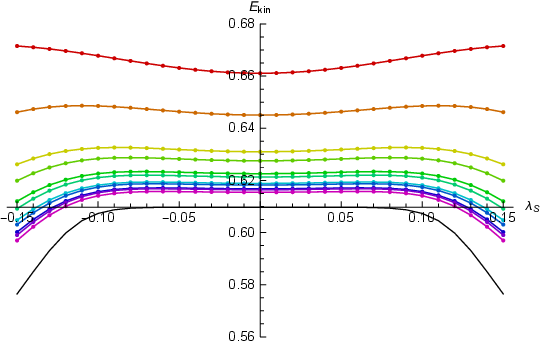}
   \end{subfigure}
   \vspace{3mm}

   \begin{subfigure}{0.45\textwidth}
   \includegraphics[width=\textwidth]{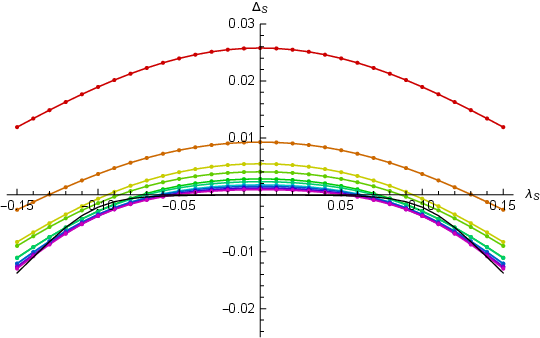}
   \end{subfigure}\qquad
   \begin{subfigure}{0.45\textwidth}
   \includegraphics[width=\textwidth]{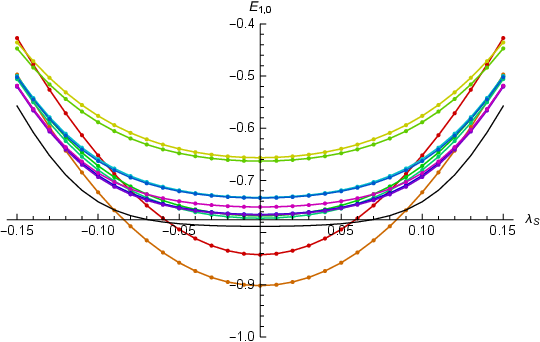}
   \end{subfigure}
   \vspace{3mm}

   \begin{subfigure}{0.45\textwidth}
   \includegraphics[width=\textwidth]{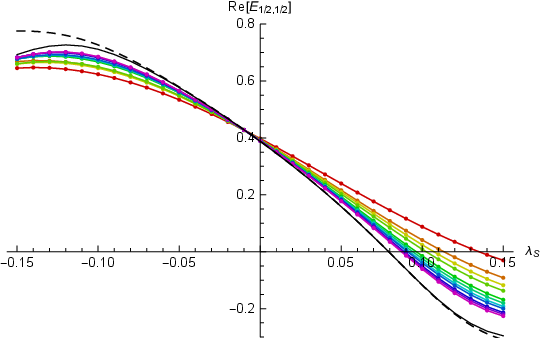}
   \end{subfigure}\qquad
   \begin{subfigure}{0.45\textwidth}
   \includegraphics[width=\textwidth]{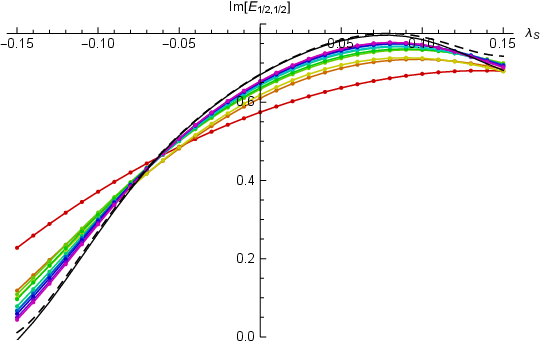}
   \end{subfigure}
   \vspace{3mm}

   \begin{subfigure}{0.45\textwidth}
   \includegraphics[width=\textwidth]{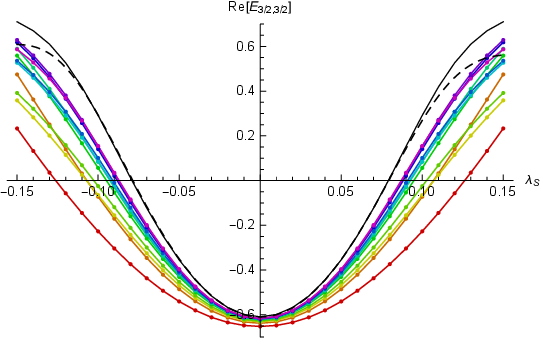}
   \end{subfigure}\qquad
   \begin{subfigure}{0.45\textwidth}
   \includegraphics[width=\textwidth]{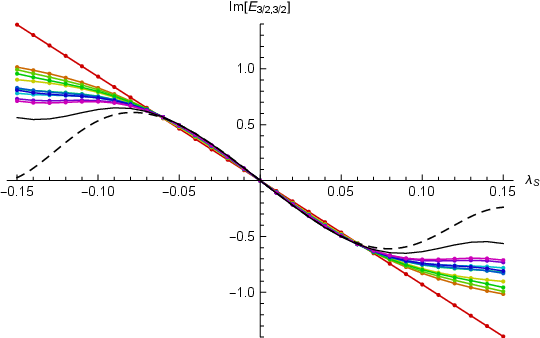}
   \end{subfigure}
   \caption{Plot of $\lS$-dependence of selected invariants for marginal deformations of the 0-brane solution in the $k=3$ model with $J=\frac{1}{2}$ boundary conditions.}
   \label{fig:k=3 0-brane}
\end{figure}

Let us begin with the 0-brane. The biggest difference compared to the 0-brane in the $k=2$ model is that the branch of solutions for $k=3$ is not fully symmetric with respect to $\lS$ and therefore we had to compute sample solutions both for positive and negative $\lS$. Have a look at figure \ref{fig:k=3 0-brane}, where we plot $\lS$-dependence of several invariants. We notice that the branch of solutions is only partially symmetric, some observables are symmetric (invariants that should not depend on $\lS$ and $E_{3/2,3/2}$) and some are not (the remaining $E_{j,m}$ invariants). The reason for the asymmetry lies in the initial geometrical configuration (which can be seen in figure \ref{fig:moduli}). The basic solution has $\theta_0=\frac{\pi}{3}$ and therefore the reflection between $\theta_0+\pi\lB$ and $\theta_0-\pi\lB$ is not a symmetry of the \SU manifold with the initial $\frac{1}{2}$-brane. In the $k=2$ case, the reference $\frac{1}{2}$-brane divides the circle in two halves and $\theta_0=\frac{\pi}{2}$, so this configuration is symmetric.

The length of the 0-brane branch for $k=3$ is again finite and it does not differ much between levels, it ends around $\lS\approx 0.27$. We evaluated sample solutions up to level 12 for $|\lS|\leq 0.15$, because, like in the $k=2$ case, the solution goes off-shell and this range of $\lS$ is enough to cover the physical part of the branch.

\begin{figure}[!t]
   \centering
   \includegraphics[width=10cm]{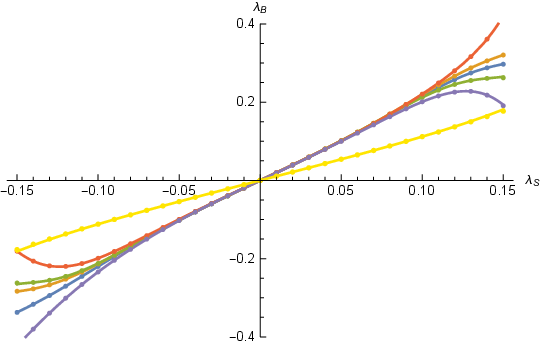}
   \caption{The relation $\lB(\lS)$ for solutions in the $k=3$ model with $J=\frac{1}{2}$ boundary conditions. The five grouped lines correspond to five independent invariants of the 0-brane branch of solutions and the remaining yellow line to the B-brane branch of solutions, which has only one invariant that depends on $\lB$, $E_{3/2,3/2}$.}
   \label{fig:k=3 lambda}
\end{figure}

Figure \ref{fig:k=3 0-brane} shows behavior of several observables of this branch of solutions. The key features are the same as for the $k=2$ solution. We observe that all invariants agree well with predictions for small $\lS$, but they deviate from the expected behavior near edges of the figures, both for positive and negative $\lS$. To find where the branch goes off-shell, we again analyzed ratios $\sigma/\sigma_{est}$ and consistency of $\lB$ (see figure \ref{fig:k=3 lambda}). We estimate that the branch becomes inconsistent at $\lS^\ast=0.10\pm 0.01$. Since some observables are symmetric (including $\Delta_S$ and the equation for the marginal field), it seems likely that the length of the on-shell part of the branch is the same both in the positive and the negative direction.

Using (\ref{lambda from elw E}) and (\ref{lambda from elw J}), we computed the boundary marginal parameter $\lB$ from Ellwood invariants. The functions $\lB(\lS)$ are plotted in figure \ref{fig:k=3 lambda}. Compared to $k=2$, there are two additional invariants ($E_{3/2,3/2}$, $E_{3/2,1/2}$), so this figure includes 5 curves. With the exception of the green curve, which corresponds to $E_{3/2,3/2}$, the functions are not antisymmetric and they do not pass through to the origin\footnote{The fits in figure \ref{fig:k=3 lambda} are done using generic polynomials in $\lS$, while we used only odd powers of $\lS$ in figure \ref{fig:k=2 lambda}. However, since the functions are almost antisymmetric for small $\lS$, the first three even parameters of the fits are small.}. However, if we make a restriction to the area $|\lS|< \lS^\ast$, where the branch is on-shell, the functions are almost antisymmetric. This suggests that the asymmetry is caused by the truncation to a finite level and it should disappear for an exact solution in the infinite level limit.

The estimated endpoint $\lS^\ast=0.10\pm 0.01$ of the on-shell part corresponds to the following value of the boundary marginal parameter:
\begin{equation}
\lB^\ast=0.21\pm 0.02.
\end{equation}
The on-shell part of the branch therefore describes about 42\% of the moduli space. Figure \ref{fig:moduli} illustrates which parts of the moduli space are covered. The on-shell part of the branch covers most of the upper half of the circle, but only a little of the lower half. This type of picture also applies for most solutions at higher $k$. Unless the initial boundary conditions are $J=k/2$, the seed solutions are concentrated only in one half of the circle in the area bounded by the initial D-brane and the other part of the circle without seeds remains mostly or completely uncovered.

\FloatBarrier
\subsection{B-brane solution}\label{sec:k=3:B-brane}
Although the main focus of this paper are solutions describing Cardy boundary states, this section to demonstrates that our approach is also applicable on solutions describing symmetry-breaking boundary states, concretely on the solution which describes a B-brane boundary state \cite{WZW B-branes} with $J=0$. From the technical point of view, there are no differences between computing deformations of a symmetry-breaking boundary state and a Cardy boundary state. We just have to start from a seed representing a B-brane instead of a 0-brane.

\begin{figure}[!t]
   \centering
   \begin{subfigure}{0.45\textwidth}
   \includegraphics[width=\textwidth]{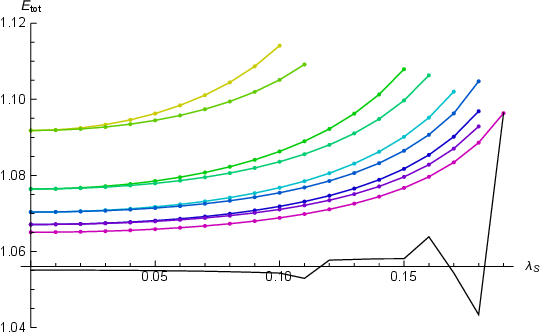}
   \end{subfigure}\qquad
   \begin{subfigure}{0.45\textwidth}
   \includegraphics[width=\textwidth]{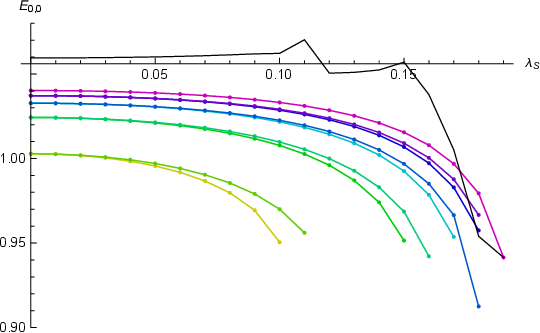}
   \end{subfigure}
   \vspace{5mm}

   \begin{subfigure}{0.45\textwidth}
   \includegraphics[width=\textwidth]{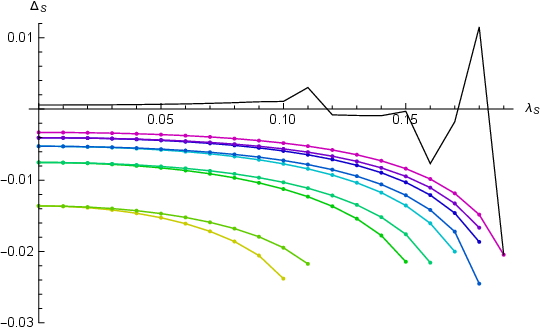}
   \end{subfigure}\qquad
   \begin{subfigure}{0.45\textwidth}
   \includegraphics[width=\textwidth]{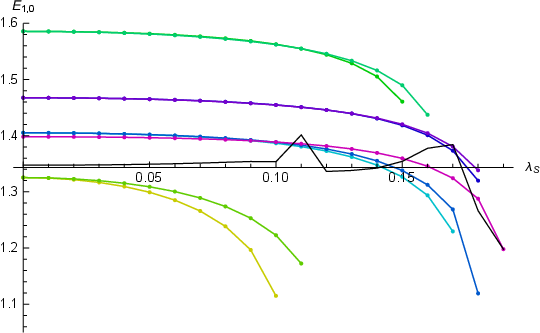}
   \end{subfigure}
   \vspace{5mm}

   \begin{subfigure}{0.45\textwidth}
   \includegraphics[width=\textwidth]{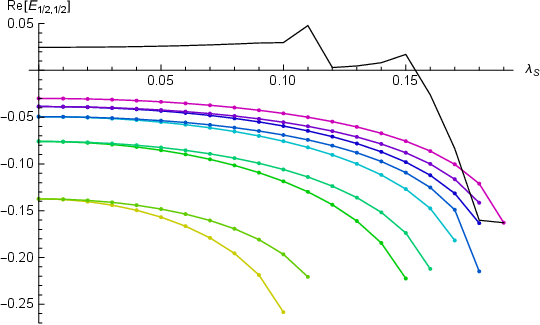}
   \end{subfigure}\qquad
   \begin{subfigure}{0.45\textwidth}
   \includegraphics[width=\textwidth]{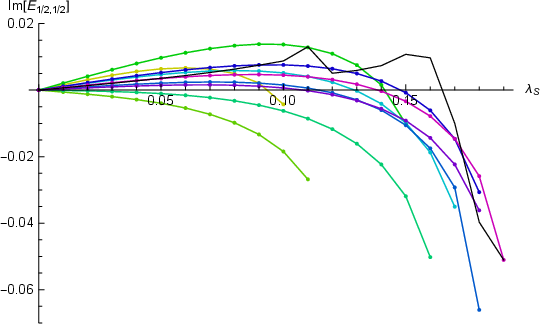}
   \end{subfigure}
   \vspace{5mm}

   \begin{subfigure}{0.45\textwidth}
   \includegraphics[width=\textwidth]{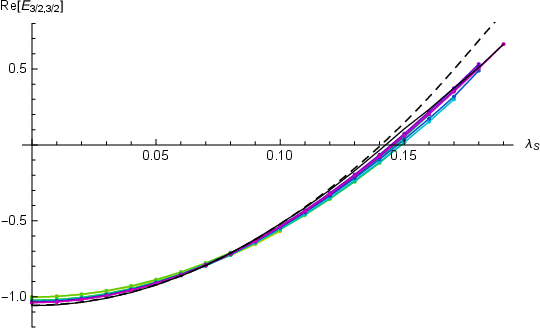}
   \end{subfigure}\qquad
   \begin{subfigure}{0.45\textwidth}
   \includegraphics[width=\textwidth]{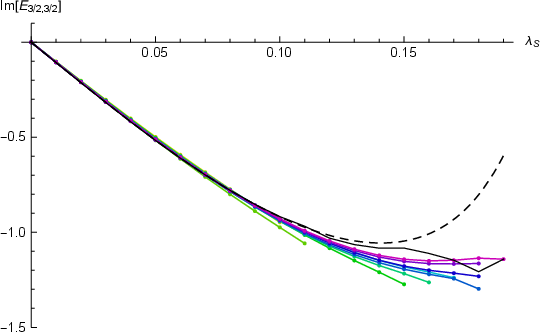}
   \end{subfigure}
   \caption{Selected invariants of the B-brane branch of solutions in the $k=3$ model with $J=\frac{1}{2}$ boundary conditions. The plots show only real data points and therefore they include only data from levels 4 to 12.}
   \label{fig:k=3 B-brane}
\end{figure}

The results regarding the B-brane solution are presented in figure \ref{fig:k=3 B-brane}, which shows plots of $\lS$-dependence of several invariants. This branch is fully (anti)symmetric with respect to $\lS=0$, so we again consider only positive $\lS$. This time, we decided to plot the whole length of the branch because we have not found evidence of the solution going off-shell. The plots show only real data points, which indicate the length of the branch at individual levels. The basic solution for $\lS=0$ is complex at levels 2 and 3 and this property transfers to the marginally deformed solution as well, so levels 2 and 3 are completely missing from the plots. However, data from higher levels also become complex for high enough $\lS$, so the data end at the latest at $\lS=0.19$ because $\lS=0.2$ solution is complex at all available levels. What happens to length of the branch at higher levels is difficult to predict, since we do not know it precise enough to make an extrapolation\footnote{To compute the length of the branch precisely, it is necessary to use bisection and repeatedly attempt to solve the equation of motion as in \cite{MarginalKMOSY}, which takes lot of CPU time.}.


Since we use only real data points for extrapolations, the black lines representing infinite level extrapolations jump up or down when the number of used data points changes. The first jump happens above $\lS=0.1$ as level 4 solution becomes complex. The precision of extrapolations decreases as more and more levels become complex and the extrapolations become completely unreliable above $\lS=0.15$ because the number of real data points becomes too low.

Most invariants of this solution (with the exception of $E_{3/2,\pm3/2}$) should be independent on $\lS$ and we observe that their extrapolations are approximately constant as long as there are enough real data points, although there is obviously a gradual loss of precision compared to the initial solution. The invariants match the predicted values quite well and there are no visible signs that the solution would go off-shell. Therefore we think that it is likely that the solution consistently describes B-brane boundary state for the whole length of the branch, although it is not possible to verify that for $\lS>0.15$ because there are too few real data points. This is an important difference compared to the 0-brane solutions above, which clearly become inconsistent long before their branches end.

Next, let us focus on the invariant $E_{3/2,3/2}$, which is the only observable that changes nontrivially with $\lS$. Marginal deformations of the B-brane boundary state associated to the $J^3$ current are
\begin{equation}\label{B-brane deformed}
\ww B,J,\lB\rra=e^{2\pi i \lB J_0^3}\ww B,J,0\rra^.
\end{equation}
The invariant $E_{3/2,3/2}$ should therefore behave in a similar way as for Cardy boundary states
\begin{equation}\label{B-brane E3/2}
E_{3/2,3/2}^{exp}=\sqrt{3} B_0^{\ 0} e^{3 \pi i \lB}.
\end{equation}
The real and imaginary part of this invariant should be proportional to cosine and sine of $3\pi \lB$. They are plotted at the bottom of figure \ref{fig:k=3 B-brane} and they are consistent with the prediction. Using an analogue of (\ref{lambda from elw E}), we computed the marginal parameter $\lB$, see figure \ref{fig:k=3 lambda}. By comparing $\lB$ of the 0-brane and the B-brane solutions, we found that there is an approximate relation between the parameters
\begin{equation}
\lB^{\rm (B-brane)}\approx \frac{1}{2}\lB^{\rm (0-brane)},
\end{equation}
which obviously holds only when both solutions are on-shell and well convergent. We will discuss relations of $\lB$ for various solutions more in section \ref{sec:comparison}.

Physical interpretation of deformations of the B-brane solution is more difficult. According to the reference \cite{WZW B-branes}, B-branes can be understood as 'fat D1-branes' and they have a 5-dimensional moduli space. Four parameters describe position of the B-brane on the 3-sphere and one parameter corresponds to Wilson line. The boundary state (6.4) from \cite{WZW B-branes} is factorized into a free boson part and a parafermion part and the $J^3$ current deformation adds a phase to the free boson part. Therefore we can interpret our branch of solutions as B-branes with Wilson line.
A solution for given $\lS$ can be further modified by action of the operators $e^{i\lambda_1 J^1_0+i\lambda_2 J^2_0}$. These deformations will cover another two dimensions of the moduli space, which will correspond to some rotations of the B-brane on the 3-sphere. Therefore we are able to cover (at least partially) 3 out of 5 dimensions of the moduli space using our OSFT approach, but it not clear how to reach the remaining two dimensions because there are no other marginal operators.

\FloatBarrier
\section{Solutions in the $k=4$ model}\label{sec:k=4}
In the reference \cite{KudrnaWZW}, we analyzed $J=1$ boundary conditions for the $k=4$ model and we found three real solutions that describe Cardy boundary states. Two of them describe 0-branes and the third describes a $\frac{1}{2}$-brane. In this section, we discuss marginal deformations of these three solutions, which we were able to evaluate up to level 11.

\subsection{0-brane solutions}\label{sec:k=4:0-brane}
In this model, there are two independent solutions for 0-branes. Each of them exists in two copies, which differ just by some signs, so the moduli space of 0-branes will be covered by marginal deformations of four seed solutions, the geometrical configuration is depicted at the bottom of figure \ref{fig:moduli}.

\begin{figure}[!]
   \centering
   \begin{subfigure}{0.45\textwidth}
   \includegraphics[width=\textwidth]{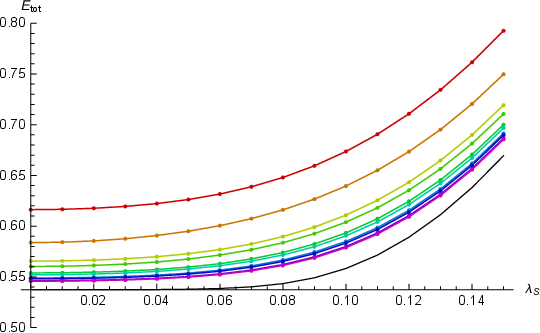}
   \end{subfigure}\qquad
   \begin{subfigure}{0.45\textwidth}
   \includegraphics[width=\textwidth]{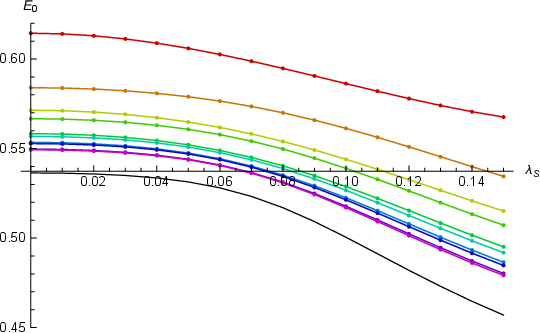}
   \end{subfigure}
   \vspace{5mm}

   \begin{subfigure}{0.45\textwidth}
   \includegraphics[width=\textwidth]{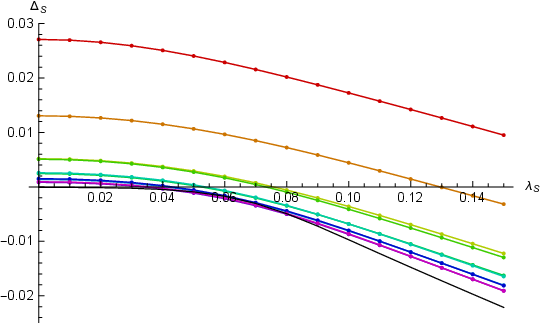}
   \end{subfigure}\qquad
   \begin{subfigure}{0.45\textwidth}
   \includegraphics[width=\textwidth]{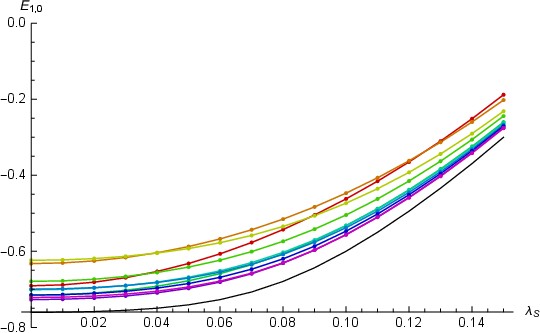}
   \end{subfigure}
   \vspace{5mm}

   \begin{subfigure}{0.45\textwidth}
   \includegraphics[width=\textwidth]{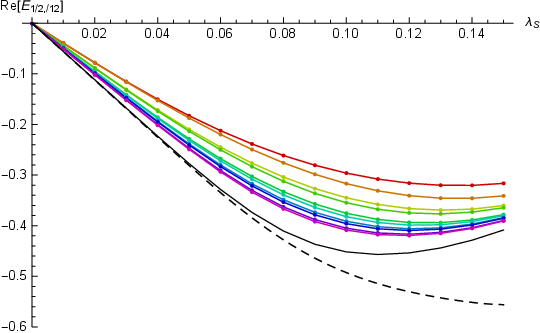}
   \end{subfigure}\qquad
   \begin{subfigure}{0.45\textwidth}
   \includegraphics[width=\textwidth]{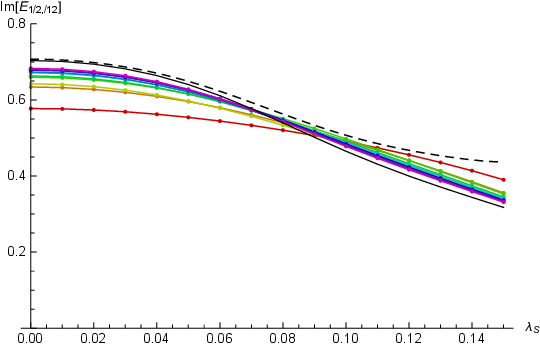}
   \end{subfigure}
   \vspace{5mm}

   \begin{subfigure}{0.45\textwidth}
   \includegraphics[width=\textwidth]{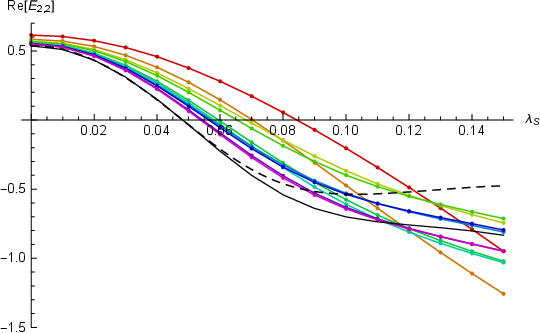}
   \end{subfigure}\qquad
   \begin{subfigure}{0.45\textwidth}
   \includegraphics[width=\textwidth]{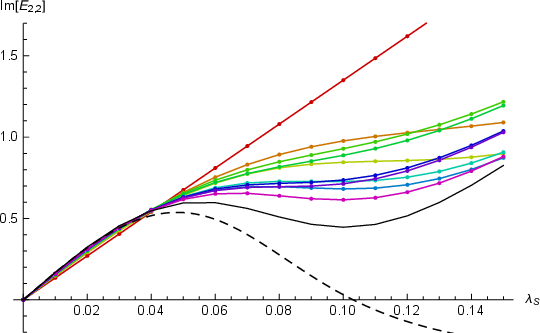}
   \end{subfigure}
   \caption{Selected observables of marginal deformations of 0-brane solution with $\theta_0=\frac{\pi}{2}$ in $k=4$ model with $J=1$ boundary conditions.}
   \label{fig:k=4 0-brane 1}
\end{figure}

The first branch of 0-brane solutions has the initial parameter $\theta_0=\frac{\pi}{2}$ and its properties are very similar to 0-branes at $k=2,3$, so we can keep the analysis short. It is symmetric with respect to $\lS=0$, so we consider once again only positive $\lS$. The branch ends around $\lS=0.23$, but it goes off-shell long before that, so the analysis will include only data up to $\lS=0.15$. Using the same style as before, some of its observables are shown in figure \ref{fig:k=4 0-brane 1}. The plots show that the branch can be interpreted as a 0-brane only in a small interval of $\lS$. We estimate that it goes off-shell already at $\lS^\ast=0.7\pm 0.1$.

Interestingly, this branch of solutions has a symmetry that connects two of its invariants:
\begin{equation}\label{symmetry}
J_{+-}=-E_{2,1}.
\end{equation}
This symmetry is different from the symmetry of real solutions described in \cite{KudrnaWZW}, which is $J_{+-}=(-1)^{2J+1}E_{k/2,k/2-1}^\ast$, because it does not include complex conjugation. However, the two symmetries are compatible for $\lS=0$ because the invariant $J_{+-}$ becomes real at this point. The origin of this symmetry is unclear.

\begin{figure}[!t]
   \centering
   \includegraphics[width=10cm]{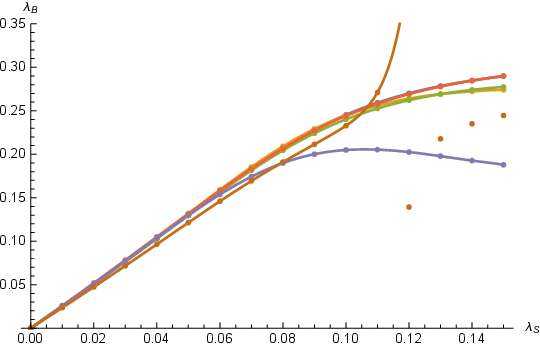}
   \caption{The relation $\lB(\lS)$ for the 0-brane solution with with $\theta_0=\frac{\pi}{2}$ in the $k=4$ model with $J=1$ boundary conditions. There is a jump in data coming from invariant $E_{2,1}$ (which are denoted by brown color) because the invariant is close to zero in a certain area, which allows a quick change of its complex phase.}
   \label{fig:k=4 lambda 0-brane 1}
   \vspace{15mm}
   \centering
   \includegraphics[width=10cm]{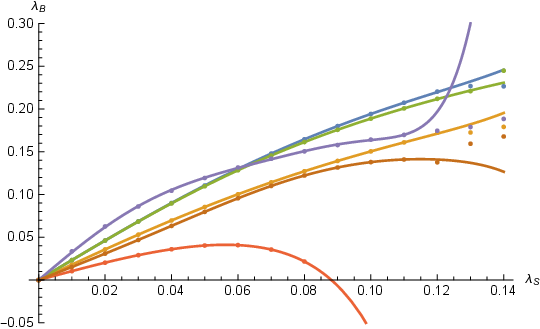}
   \caption{The relation $\lB(\lS)$ for the 0-brane solution with with $\theta_0=0$ in the $k=4$ model with $J=1$ boundary conditions. The polynomial fits are based only on data with  $\lS\leq 0.11$.}
   \label{fig:k=4 lambda 0-brane 2}
\end{figure}

In figure \ref{fig:k=4 lambda 0-brane 1}, we plot values of the boundary marginal parameter $\lB$ computed from the independent invariants. One set of points is missing from the plot because the results from $J_{+-}$ and $E_{2,1}$ are identical due to the symmetry. The figure suggests that the end of the on-shell part of the branch corresponds to $\lB^\ast=0.18\pm 0.02$.

\begin{figure}[!]
   \centering
   \begin{subfigure}{0.45\textwidth}
   \includegraphics[width=\textwidth]{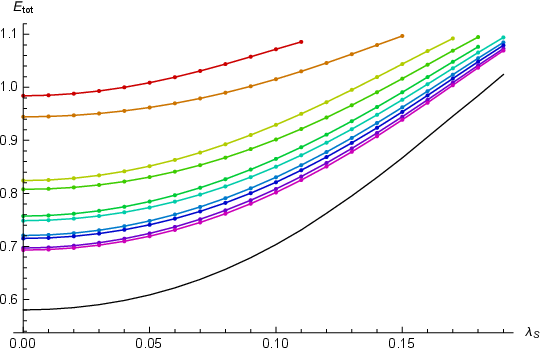}
   \end{subfigure}\qquad
   \begin{subfigure}{0.45\textwidth}
   \includegraphics[width=\textwidth]{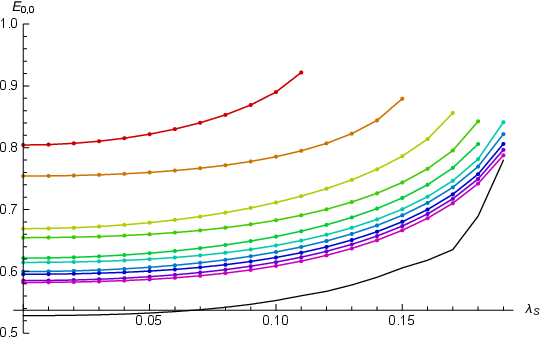}
   \end{subfigure}
   \vspace{5mm}

   \begin{subfigure}{0.45\textwidth}
   \includegraphics[width=\textwidth]{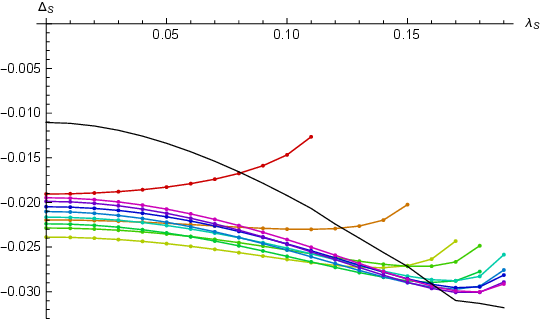}
   \end{subfigure}\qquad
   \begin{subfigure}{0.45\textwidth}
   \includegraphics[width=\textwidth]{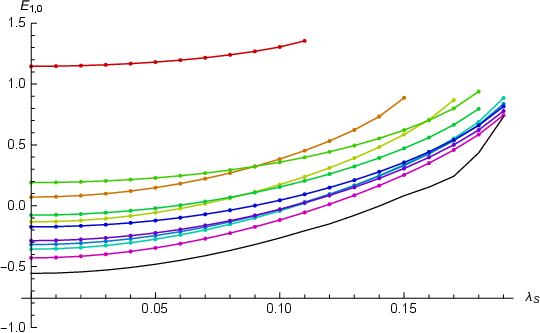}
   \end{subfigure}
   \vspace{5mm}

   \begin{subfigure}{0.45\textwidth}
   \includegraphics[width=\textwidth]{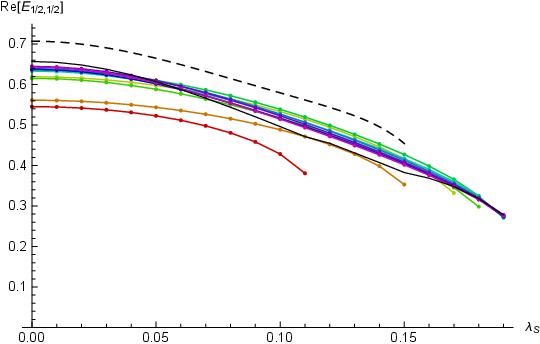}
   \end{subfigure}\qquad
   \begin{subfigure}{0.45\textwidth}
   \includegraphics[width=\textwidth]{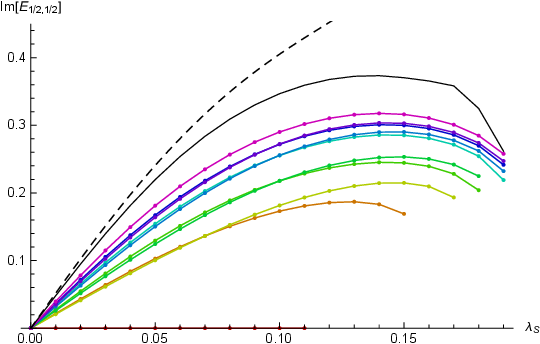}
   \end{subfigure}
   \vspace{5mm}

   \begin{subfigure}{0.45\textwidth}
   \includegraphics[width=\textwidth]{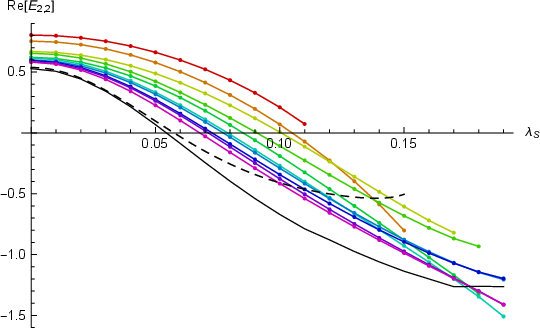}
   \end{subfigure}\qquad
   \begin{subfigure}{0.45\textwidth}
   \includegraphics[width=\textwidth]{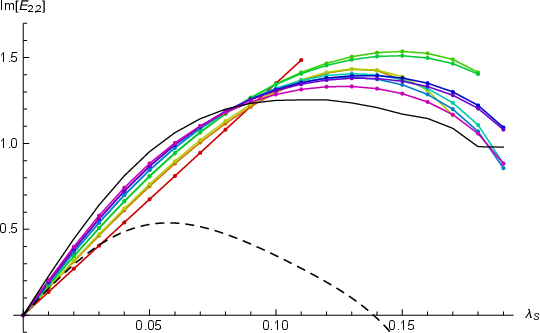}
   \end{subfigure}
   \caption{Selected observables of marginal deformations of the 0-brane solution with $\theta_0=0$ in the $k=4$ model with $J=1$ boundary conditions.}
   \label{fig:k=4 0-brane 2}
\end{figure}

The second 0-brane branch of solutions with $\theta_0=0$ has different properties and its most important characteristic is, unfortunately, its low precision. The data quickly change with level and our extrapolation method does not work very well, so the precision is at least by one order worse than for the other 0-brane branch with $\theta_0=\frac{\pi}{2}$. The identification of the seed solution in \cite{KudrnaWZW} was actually based more on the fact that it fits a generic pattern satisfied by Cardy brane solutions than on a precise agreement with the expected boundary state.

Nevertheless, we can do marginal deformations of this solution. Tracking the solution is a bit more difficult because Newton's method fails for $\lS>0.1$ if use the level truncation scheme in the traditional way and improve solutions level by level. Therefore we modify our approach a bit and we take the seed for Newton's method to be the solution from the same level with lower value of $\lS$ instead of a solution from lower level. This approach works well and we can track the branch till its end. The branch of solutions becomes complex slightly below $\lS=0.2$ at most levels. To demonstrate properties of this branch of solutions, we plot $\lS$-dependence of several invariants in figure \ref{fig:k=4 0-brane 2}.

As stated above, the basic solution has a poor precision and it decreases even further as we turn on the marginal field. Therefore we are unable to make a definite claim whether (and possible where) it goes off-shell like the other 0-brane solutions. We are not even entirely sure that the basic solution is consistent, so there is even the possibility of the entire branch being off-shell. However, a more likely option seems to be that it is consistent in a small interval of $\lS$ before going off-shell.
For illustration, let us have a look at some of the invariants. The energy, which should converge to 0.537285, starts at 0.58 for $\lS=0$ and it grows approximately towards 1. The invariant $E_{0,0}$ is more precise for small $\lS$, but it also grows far above the expected value. The $\lS$-dependent invariants have the right tendency, but their actual values are often quite off. Consider, for example, the invariant $E_{2,2}$, which is plotted at the bottom of figure \ref{fig:k=4 0-brane 2}. Absolute value of this invariant should be also 0.537285, but the actual values surpass 1.5 for some $\lS$.

In summary, the solution very roughly agrees with the predictions for small $\lS$, but it is unlikely that it is consistent for large $\lS$. However, the solution evolves rapidly with level and if we had data from few more levels, we expect that the extrapolations would change significantly, although it is not certain whether they would be closer to the expected components of a 0-brane boundary state or not.

Figure \ref{fig:k=4 lambda 0-brane 2} shows the values of $\lB$ extracted from the solution. This solution also has the symmetry (\ref{symmetry}), so data corresponding to $J_{+-}$ are missing. Unlike for the previous cases, the curves deviate from each other right from the beginning, which is a result of poor precision of the extrapolations. The closest match between the curves is between the blue and green ones, which correspond to invariants $E_{1/2,1/2}$ and $E_{3/2,3/2}$ respectively. Since we do not know how much of the branch is on-shell and which of the curves is closest to the actual relation $\lB(\lS)$, we are not able to tell how much of the moduli space is covered by this branch of solutions. Therefore the part of figure \ref{fig:moduli} that corresponds to this branch is only schematic and it is not based on a concrete numeric result.

\FloatBarrier
\subsection{$\frac{1}{2}$-brane solution}\label{sec:k=4:1/2-brane}
\begin{figure}[!]
   \centering
   \begin{subfigure}{0.45\textwidth}
   \includegraphics[width=\textwidth]{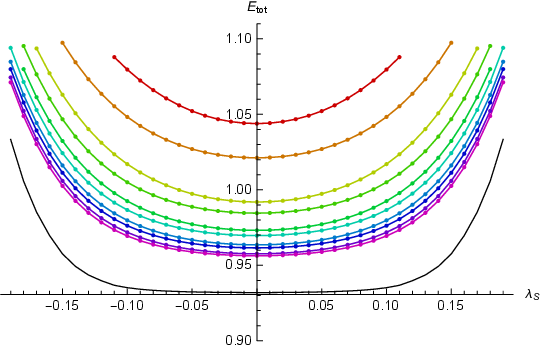}
   \end{subfigure}\qquad
   \begin{subfigure}{0.45\textwidth}
   \includegraphics[width=\textwidth]{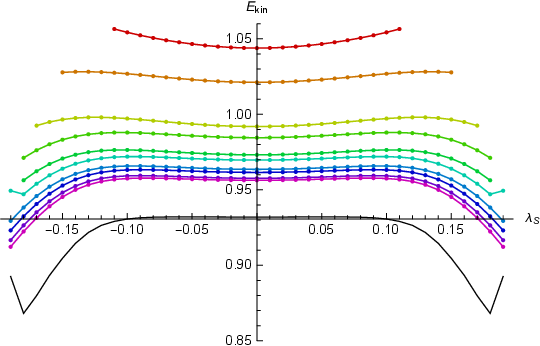}
   \end{subfigure}
   \vspace{5mm}

   \begin{subfigure}{0.45\textwidth}
   \includegraphics[width=\textwidth]{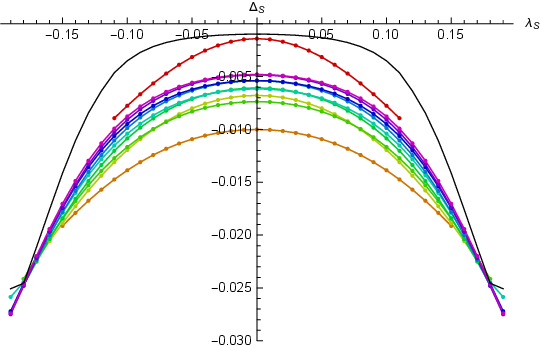}
   \end{subfigure}\qquad
   \begin{subfigure}{0.45\textwidth}
   \includegraphics[width=\textwidth]{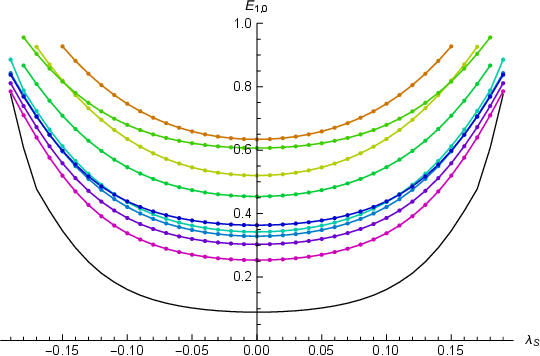}
   \end{subfigure}
   \vspace{5mm}

   \begin{subfigure}{0.45\textwidth}
   \includegraphics[width=\textwidth]{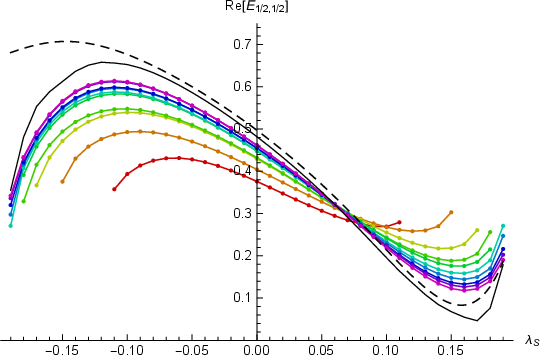}
   \end{subfigure}\qquad
   \begin{subfigure}{0.45\textwidth}
   \includegraphics[width=\textwidth]{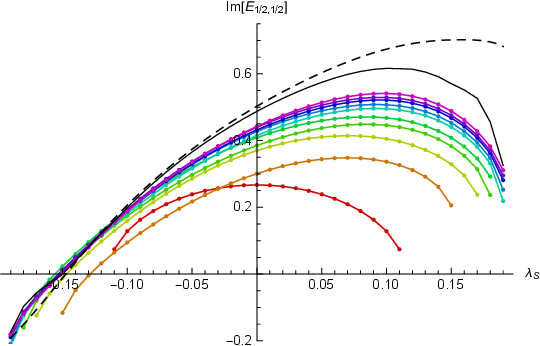}
   \end{subfigure}
   \vspace{5mm}

   \begin{subfigure}{0.45\textwidth}
   \includegraphics[width=\textwidth]{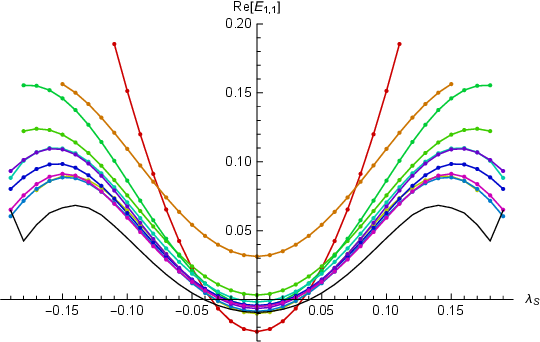}
   \end{subfigure}\qquad
   \begin{subfigure}{0.45\textwidth}
   \includegraphics[width=\textwidth]{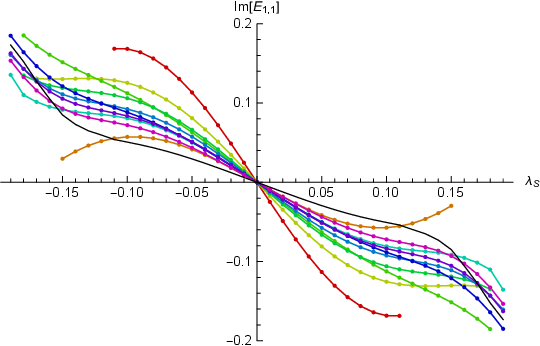}
   \end{subfigure}
   \caption{Selected observables of marginal deformations of the $\frac{1}{2}$-brane solution in the $k=4$ model with $J=1$ boundary conditions.}
   \label{fig:k=4 1/2-brane}
\end{figure}

Finally, let us discuss marginal deformations of the $\frac{1}{2}$-brane solution with $\theta_0=\frac{\pi}{4}$. This branch of solutions is only partially symmetric (similarly to the 0-brane for $k=3$), see figure \ref{fig:k=4 1/2-brane}, so we had to compute sample solutions both for positive and negative $\lS$. The symmetric quantities are those that should not depend on $\lS$ and invariants $E_{1,1}$ and $E_{2,2}$, while the remaining invariants are asymmetric.

In this case, we decided to show the full length of the branch because most of it seems to be on-shell. The length of the branch is around $0.19$ at higher levels, but it is quite short at level 2. This is a slight complication because the level 2 branch probably ends before the solution goes off-shell and there are jumps of extrapolations in the critical area.

As usual, we plot several invariants of this branch of solutions in figure \ref{fig:k=4 1/2-brane} to illustrate its behavior. The precision is somewhere between the two 0-brane branches, which is not very surprising because the same holds for the seed solutions.
This solution does not have the symmetry (\ref{symmetry}), which means it has the maximal number of independent invariants, but there is a connection between $J_{+-}$ and $E_{2,1}$. We noticed that these invariants are related when we switch the sign of $\lS$:
\begin{equation}
J_{+-}(\lS)=-E_{2,1}^\ast (-\lS).
\end{equation}

We estimate that the on-shell part of the branch ends approximately at
\begin{equation}
\lS^\ast = 0.12\pm 0.02,
\end{equation}
where we consider a larger uncertainty than for other solutions due to jumps of infinite level extrapolations caused by end of the level 2 branch. They are not as visible as for the B-brane solution in figure \ref{fig:k=3 B-brane}, but they affect the ratios $\sigma/\sigma_{est}$.

\begin{figure}[!t]
   \centering
   \includegraphics[width=10cm]{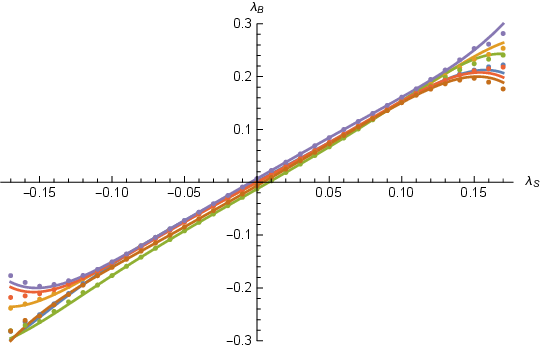}
   \caption{The relation $\lB(\lS)$ of the $\frac{1}{2}$-brane solution in the $k=4$ model with $J=1$ boundary conditions. The polynomial fits are done using only data with $|\lS|\leq 0.11$ because there is a jump in extrapolations as level 2 branch ends.}
   \label{fig:k=4 lambda 1/2-brane}
\end{figure}

The relation between the two marginal parameters is plotted in figure \ref{fig:k=4 lambda 1/2-brane}. This time, the plot does not include the invariant $E_{1,1}$, which should converge to zero. From the plot, we read off that the end of the on-shell part of the branch corresponds to
\begin{equation}
\lB^\ast = 0.19\pm 0.3.
\end{equation}
That correspond approximately to 37\% of the moduli space.

\FloatBarrier
\section{Comparison of $\lB$ for different solutions}\label{sec:comparison}
In the final section, we would like to do a comparison of solutions on different backgrounds. We will focus on the relation $\lB(\lS)$ because $\lB$ is a common parameter of all branches of solutions and it has a clear physical meaning.

First, let us discuss normalization of the marginal parameters. There is a freedom in definition of the marginal parameters because both can be multiplied by a constant. In this paper, we define $\lS$ as the string field coefficient in front of the marginal field (\ref{lambdaS def}) and $\lB$ using (\ref{lambdaB def}), which leads to the parameterization of the angle $\theta=\theta_0+\pi \lB$. With this normalization, the two parameters match in the first order for the standard marginal solution,
\begin{equation}\label{lambda standard}
\lB(\lS)=\lS+O(\lS^3),
\end{equation}
which can be seen from the perturbative expansion of the solution:
\begin{equation}
\Psi(\lS)=\lS J_{-1}^3 c_1 |0,0\ra+O(\lS^3) .
\end{equation}
The relation (\ref{lambda standard}) does not depend on the level $k$ or the initial boundary conditions $J$.
However, the marginal state $J_{-1}^3c_1|0,0\ra$ has norm $k$, so if we wanted the marginal field to have unit normalization, we would have to scale $\lS$ by a factor of $\sqrt{k}$. In order to preserve (\ref{lambda standard}), $\lB$ would have to be multiplied by the same factor as well.

When it comes to the families of solutions studied in this work, the relation between the two marginal parameters is also linear for small $\lS$, but it does not satisfy (\ref{lambda standard}) because there is a generic proportionality constant.

After analyzing our data, we realized that it is interesting to compare functions $\lB(\sqrt{k}\lS)$, which seem to be related. This is somewhat surprising because only one of the two marginal parameters is multiplied by $\sqrt{k}$ to match the unit normalization and it leads to change of the relation (\ref{lambda standard}) for the standard marginal solution. $\lB$ is proportional to $\Delta\theta$, so can also view this relation as dependence of the angle $\theta$ on the normalized OSFT marginal parameter.

\begin{figure}[!t]
   \centering
   \includegraphics[width=12cm]{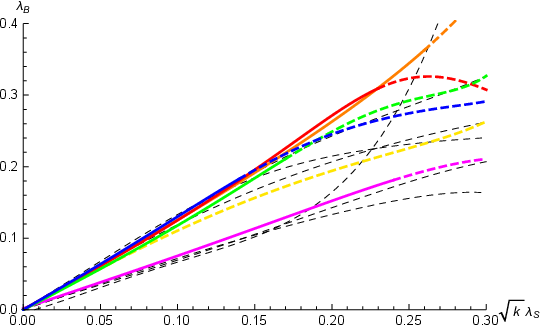}
   \caption{The relation between the marginal parameters $\sqrt{k}\lS$ and $\lB$ for various solutions. In case of the B-brane solution (orange), we plot $2\lB$.}
   \label{fig:lambda}
\end{figure}

The functions $\lB(\sqrt{k}\lS)$ are plotted in figure \ref{fig:lambda}. The results include the solutions that we analyzed in the previous sections and few more examples of solutions from $k=5,6$ models. The solutions from the previous sections are represented by colorful lines that are solid when we consider the functions to be reliable (the corresponding solutions are on-shell and infinite level extrapolations are trustworthy), while the dashed parts of lines denote unreliable continuation of the fits. $k=5,6$ solutions are represented by thinner black dashed lines. The plotted functions are given by polynomial fits of $\lB$ computed from the invariant $E_{1/2,1/2}$, with the exception of the B-brane branch of solutions, which uses $E_{3/2,3/2}$.

The red, green and blue curves represent 0-brane branches of solutions in $k=2,3,4$ models respectively (with $\theta_0=\frac{\pi}{2}$ in the $k=4$ case). We notice that the solid parts of the curves, which denote on-shell parts of the branches, are close to each other.
This suggests that 0-brane branches of solutions have the same first coefficient of the expansion of $\lB(\sqrt{k}\lS)$ in the infinite level limit:
\begin{equation}
\lB(\lS)=a_1 \sqrt{k}\lS+O(\lS^3),
\end{equation}
where $a_1\approx 1.2$. The precision of our results is not good enough to claim that with certainty, but it is not probable that the curves would be so close to each other if they were unrelated. It is also possible that the exact solutions for 0-branes share not only the coefficient $a_1$, but the whole function $\lB(\sqrt{k}\lS)$. Our data are however not precise enough to make a reliable comparison of third order or higher coefficients.

Next, there is a function that corresponds to the branch of 0-brane solutions in the $k=4$ model with $\theta_0=0$, which is denoted by yellow color. The precision of this branch of solutions is very low and the fit is not very reliable, but for small $\lS$, the curve is not very far from the other 0-brane branches, so it is plausible that it also has the same coefficient $a_1$.

The magenta line represents the $\frac{1}{2}$-brane branch of solutions in the $k=4$ model. We observe that this branch clearly has a different relation between the marginal parameters from 0-branes. The first coefficient in the expansion is approximately $a_1\approx 0.7$ and it does not seem to be related to $a_1$ for 0-branes in a way we recognize (like by a simple rational number).

The orange line corresponds to the B-brane branch of solutions in the $k=3$ model. It matches very nicely the 0-brane curves, but it represents two times $\lB(\sqrt{k}\lS)$. Therefore we have approximately
\begin{equation}
\lB(\sqrt{k}\lS)^{\rm (B-brane)}\approx\frac{1}{2}\lB(\sqrt{k}\lS)^{\rm (0-brane)}.
\end{equation}
The agreement may again include either just the first order coefficient or the whole function.

Finally, we there are also some 0-brane and $\frac{1}{2}$-brane solutions from $k=5,6$ models, which we added to have more examples. They are denoted by dashed black lines. These solutions were computed to lower levels and they have typically lower precision, but we observe that, for enough small $\lS$, all of them are close either to the cluster of lines representing 0-branes or to the magenta line representing $\frac{1}{2}$-brane.

The results from this section lead us to an interesting conjecture. They suggest that branches of marginal solutions which are not connected to the perturbative vacuum are divided into several groups based on the label $J$ of the final boundary state (the initial boundary condition does not seem to matter as long as it is different from the final one). We conjecture that in the infinite level limit, where the solutions should be exact, branches of solutions from each group share either the first coefficient in the expansion of $\lB(\sqrt{k}\lS)$ or the whole function.

Interestingly, the B-brane branch of solutions may be also related to the 0-brane group because its boundary marginal parameter seems to differ only by a factor of $\frac{1}{2}$. A possible reason why there is a relation between the B-brane and 0-branes marginal parameters is that B-branes also have a half-integer label $J$ \cite{WZW B-branes}, which equals 0 for the branch we study. It is possible that B-brane solutions with $J=\frac{1}{2}$ may have $\lB$ related to $\frac{1}{2}$-branes, but we have not found such solutions yet.

Finally, let us have a quick look at the lengths of branches of 0-brane solutions. The $k=2$ branch goes off-shell approximately at $\sqrt{k}\lS\approx 0.23$, the $k=3$ branch around $\sqrt{k}\lS\approx 0.17$ and the $k=4$ branch around $\sqrt{k}\lS\approx 0.14$. Therefore the lengths of on-shell parts of branches representing 0-branes get shorter with increasing level $k$ and they also describe a smaller interval of $\lB$. This property reflects the structure of seed solutions, which lie on a lattice of $k$ points. Therefore the distances between them get shorter with increasing $k$ and it is natural that the interval of $\lB$ described by one branch of solutions gradually gets smaller and smaller. The decreasing length of branches also means that the whole set of marginal solutions will generally cover a smaller percentage of the moduli space as $k$ increases. Especially for initial boundary conditions that offer only a small number of seed solutions, like $J=1/2$.

\section*{Acknowledgements}

\noindent

Our work has been funded by the Grant Agency of Czech Republic under the grant EXPRO 20-25775X.
Computational resources were provided by the e-INFRA CZ project (ID:90254), supported by the Ministry of Education, Youth and Sports of the Czech Republic.

\FloatBarrier

\end{document}